\documentclass[11pt,a4paper]{article}
\usepackage{jcappub}
\usepackage[latin9]{inputenc}
\usepackage{textcomp}
\usepackage{amsmath}
\usepackage{amssymb}
\usepackage{graphicx}
\usepackage[normalem]{ulem}

\makeatletter

\usepackage{euscript}\usepackage{epsfig}\usepackage{amsfonts}\usepackage{latexsym}\usepackage{color}

\newcommand{\be}[1]{\begin{equation}\label{#1}}
\newcommand{\ee}{\end{equation}}
\newcommand{\ba}[1]{\begin{eqnarray}\label{#1}}
\newcommand{\ea}{\end{eqnarray}}
\newcommand{\rf}[1]{(\ref{#1})}
\newcommand{\nn}{\nonumber}

\newcommand{\dt}{\,\partial}
\newcommand{\gdet}{\sqrt{|g|}\,}

\newcommand{\bphi}{\phi_c}
\newcommand{\pphi}{\varphi}

\newcommand{\cH}{\mathcal{H} }

\makeatother

\begin{document}

\title{$K$-essence model from the mechanical approach point of view: coupled scalar field and the late cosmic acceleration}

\author{Mariam Bouhmadi-L\'opez$^{1,2,3,4}$,}
\author{K.~Sravan~Kumar$^{1,2}$,}
\author{Jo\~ao~Marto$^{1,2}$,}
\author{Jo\~ao~Morais$^{3}$,}
\author{and Alexander Zhuk$^{5}$}

\affiliation{$^{1}$Departamento de F\'{i}sica, Universidade da Beira Interior, 6200 Covilh\~a, Portugal \\}

\affiliation{$^{2}$Centro de Matem\'atica e Aplica\c{c}\~oes da Universidade da Beira Interior (CMA-UBI),\\ 6200 Covilh\~{a}, Portugal\\}

\affiliation{$^{3}$Department of Theoretical Physics, University of the Basque Country
UPV/EHU,\\ P.O. Box 644, 48080 Bilbao, Spain\\}

\affiliation{$^{4}$IKERBASQUE, Basque Foundation for Science, 48011 Bilbao, Spain\\}

\affiliation{$^{5}$Astronomical Observatory, Odessa National University, Street Dvoryanskaya 2,\\ Odessa 65082, Ukraine\\}

\emailAdd{mbl@ubi.pt On leave of absence from UPV and IKERBASQUE}
\emailAdd{sravan@ubi.pt}
\emailAdd{jmarto@ubi.pt}
\emailAdd{jviegas001@ikasle.ehu.eus}
\emailAdd{ai.zhuk2@gmail.com}

\abstract{In this paper, we consider the Universe at the late stage of its evolution and deep inside the cell of uniformity. At these scales, we can consider the Universe to be filled with dust-like matter in the form of discretely distributed galaxies, a $K$-essence scalar field, playing the role of dark energy, and radiation as matter sources. We investigate such a Universe in the mechanical approach. This means that the peculiar velocities of the inhomogeneities (in the form of galaxies) as well as the fluctuations of the other perfect fluids are non-relativistic. Such fluids are designated as coupled because they are concentrated around the inhomogeneities. In the present paper, we investigate the conditions under which the $K$-essence scalar field with the most general form for its action can become coupled. We investigate at the background level three particular examples of the $K$-essence models:
(i) the pure kinetic $K$-essence field, (ii) a $K$-essence with a constant speed of sound and (iii) the $K$-essence model with the Lagrangian $bX+cX^2-V(\phi)$. We demonstrate that if the $K$-essence is coupled, all these $K$-essence models take the form of multicomponent perfect fluids where one of the component is the cosmological constant. Therefore, they can provide the late-time cosmic acceleration and be simultaneously compatible with the mechanical approach. }

\maketitle

\flushbottom

\

\section{Introduction}

\

The late-time cosmic acceleration detected a bit more than a decade ago \cite{SN1,SN2} is still one of the biggest challenge of modern physics and cosmology.
A number of alternative theories trying to explain this big puzzle were proposed. Most of these proposals are coined with  the name dark energy (DE). The models using scalar fields to explain DE are amongst the most popular ones. These are called quintessence \cite{quintess1,quintess1a,quintess2,quintess3} whenever the equation of state (EoS) parameter satisfies $-1< w <0$, phantom \cite{phantom1,phantom2} when $ w <-1$, and quintom \cite{quintom} if there is $w=-1$ crossing. One important and interesting feature of such models is that they can have a dynamical EoS which may solve the coincidence problem \cite{Dolgov}. However, there is also the possibility to construct models with constant $w$. This imposes severe restrictions on the form of the scalar field potential \cite{zhuk1996,ZBG,ZBG2}.

The $K$-essence scalar field models are one of the most general forms of scalar field theories. Here, the Lagrangian is a general function of the scalar field $\phi$ and its kinetic term $X$. Originally, such models were propose to study the inflationary era \cite{ArmendarizPicon:1999rj,Garriga:1999vw}. However, after the discovery of the late-time cosmic acceleration, it was realized that these types of models are also suitable to explain this phenomena (see e.g.  the book \cite{book} and references therein).

In the present paper, we study the $K$-essence model at  the background and perturbed levels. The study at the perturbed level is carried for the late-time Universe and deep inside the cell of uniformity\footnote{\label{foot1}For an estimation of the dimension of the cell of uniformity (the scale of homogeneity)  for the gravitational interaction  at large distances, see \cite{Ein1}.}.  In this regime, the Universe is highly inhomogeneous. We can clearly observe inhomogeneities in the form of galaxies and group of galaxies which have non-relativistic peculiar velocities. Under such conditions, the mechanical approach \cite{Eingorn:2012jm,EZcosm2} is an adequate tool to study  the cosmological scalar perturbations. Moreover, it enables us to get the gravitational potential and to consider the motions of galaxies \cite{EKZ2}.  On top of  the inhomogeneities, we include matter (in our case this is of the form of a $K$-essence scalar field) in  a very specific form such that their fluctuations have also non-relativistic peculiar velocities. This usually happens when these fluctuations are concentrated around the inhomogeneities (e.g. around the galaxies). In this sense, these fluids are coupled to the inhomogeneities \cite{coupled}. Such fluids were considered in our previous papers \cite{Burgazli:2013qy,ENZ1,Laslo2,Bouhmadi-Lopez:2015oxa,Burgazli:2015mzm}. On the other hand, we have shown \cite{CPL} that fluids with a linear EoS (e.g. the Chevallier-Polarski-Linder model) cannot be coupled.

On our previous paper \cite{Burgazli:2015mzm}, we have considered a canonical scalar field and demonstrated
that this scalar field can be coupled. At the background level,  such a coupled scalar field is being considered as a one-component perfect fluid that has a constant EoS parameter $w=-1/3$, unsuitable, therefore, to explain the late-time acceleration for the Universe. In the present paper, we generalize our analysis to the case of $K$-essence scalar fields. First, we define the conditions under which the $K$-essence scalar field (with the most general $K$-essence action) is coupled. Then, at the background level we consider particular examples of the $K$-essence fields  in the case where they satisfy the coupling conditions. For all these examples, we show that the $K$-essence scalar field takes the form of a multicomponent perfect fluid where one of the component represents a cosmological constant. It happens even in the case of a pure kinetic model where a potential energy term is absent. Therefore, such fluids can provide the late-time cosmic acceleration and at the same time be compatible with the mechanical approach. The last two sentences summarize the main result of our present work.

The paper is structured as follows. The background equations are given in Sec.~\ref{sec:2}. In Sec.~\ref{sec:3}, we investigate within the mechanical approach the scalar perturbations for the considered cosmological model with the most general action for the $K$-essence scalar field, and define the conditions under which the scalar field satisfies the coupling conditions. In Sec.~\ref{sec:4}, we investigate at the background level three particular examples of the $K$-essence models: (i) the pure kinetic $K$-essence field, (ii) a $K$-essence with a constant speed of sound and (iii) the $K$-essence model with the Lagrangian $bX+cX^2-V(\phi)$. We demonstrate that if they are coupled, all these $K$-essence
scalar fields take the form of multicomponent perfect fluids where one of the component is the cosmological constant. The main results are summarised in the concluding Sec.~\ref{sec:5}.

\section{\label{sec:2}Background equations}

\

\setcounter{equation}{0}

The $K$-essence field action has a general form \cite{ArmendarizPicon:1999rj,Garriga:1999vw,book}:
\be{1.1}
S_{P}=\int \mathrm{d}^{4}{x}\gdet\,\mathcal{L}_{P}=\int \mathrm{d}^{4}{x}\gdet\, P\left(X,\phi\right),\qquad X=\frac{1}{2}g^{\mu\nu}\partial_{\mu}\phi\partial_{\nu}\phi\, ,
\ee
where $P$ is an arbitrary function of $X$ and $\phi$.
The field equation is obtained by varying this action with regards to $\phi$:
\be{1.2}
\frac{1}{\gdet}\dt_{\mu}\left(\gdet g^{\mu\nu}P_{X}\dt_{\nu}\phi\right)-P_{\phi}=0\, .
\ee
Henceforth, and whenever there is no risk of confusion in the notation, the subscripts $X$- and $\phi$ indicate derivatives with respect to $X$ and $\phi$, respectively, i.e., $f_X \equiv (\partial f/\partial X)$ and $f_\phi \equiv (\partial f/\partial \phi)$.

The $K$-essence energy-momentum tensor is \cite{ArmendarizPicon:1999rj,Garriga:1999vw,book}:
\ba{1.3}
T_{\mu\nu} &=&2\frac{\delta\mathcal{L}_{P}}{\delta g^{\mu\nu}}-g_{\mu\nu}\mathcal{L}_{P}
=2{P_{X}}\frac{\delta X}{\delta g^{\mu\nu}}-g_{\mu\nu}P\nonumber \\
& =&{P_{X}}\dt_{\mu}\phi\dt_{\nu}\phi-g_{\mu\nu}P
=2X{P_{X}}u_{\mu}u_{\nu}-g_{\mu\nu}P\, ,
\ea
where the 4-momentum vector $u_{\mu}\equiv\dt_{\mu}\phi/\left(2X\right)^{1/2}$ is defined
in \cite{ArmendarizPicon:1999rj,Garriga:1999vw,book} in analogy with the
case of a minimally coupled scalar field \cite{Malik:2008im}. Comparison of \rf{1.3} with the perfect fluid formulation
\be{1.4}
	T_{\mu\nu}=\left(\varepsilon+P\right)u_{\mu}u_{\nu}-g_{\mu\nu}P
\ee
leads to the expression for the energy density of the $K$-essence field \cite{ArmendarizPicon:1999rj,Garriga:1999vw,book}
\be{1.5}
	\varepsilon=2X{P_X}-P
	\, .
\ee
Using this definition we can express the EoS, $w$, and the squared speed of sound, $c_s^2$, as \cite{Garriga:1999vw,book}:
\ba{1.6}
	&{}&w \equiv \frac{P}{\varepsilon}
	= \frac{P}{2X{P_X}-P}
	= \left[2\frac{\dt \ln P}{\dt \ln X}-1\right]^{-1}
	\, ,\\
	\label{1.7}
	&{}&c_s^2 \equiv \frac{{P_X}}{\varepsilon_X}
	= \frac{{P_X}}{2X{P_{XX}}+{P_X}}
	= \left[2\frac{\dt \ln {P_X}}{\dt \ln X}+1\right]^{-1}
	\, .
\ea

In a Friedmann-Lema\^itre-Robertson-Walker (FLRW) Universe, we will stick to this symmetry on the rest of this section, the conservation of the energy density of the $K$-essence reads
\be{1.8}
	\left(\bar\varepsilon\right)'+3\cH\left(\bar\varepsilon+\bar P\right)=0
	\, .
\ee
Hereafter, an overline denotes quantities related to the background FLRW model. Obviously, they depend on time but not on spatial coordinates. We can write Eq. \rf{1.8} as
\be{1.9}
\bar\varepsilon_X {\bar X}'+\bar\varepsilon_\phi \bphi'+6\cH \bar X{\bar P_X}=0
	 \, ,
\ee
where $\bar\varepsilon_X=\frac{\partial\varepsilon}{\partial X} (\bar X,\phi_c), \bar\varepsilon_\phi=\frac{\partial\varepsilon}{\partial\phi}(\bar X,\phi_c),
\bar P_X=\frac{\partial P}{\partial X} (\bar X,\phi_c),$ etc, and $\phi_c = \phi_c(\eta)$ is the background value of scalar field. Superscript prime denotes derivative with respect to the conformal time $\eta$ related to the synchronous time $t$ as $dt=ad\eta$, where $a$ is the scale factor of the Universe. Finally, $\mathcal{H}$ is the conformal Hubble parameter, i.e. $\mathcal{H}\equiv a'/a$.

If we take into account the definition of the squared speed of sound \rf{1.7}  and that in a FLRW Universe we have
\be{1.10}
	\bar X=\frac{1}{2a^{2}}\left(\bphi'\right)^{2}
	\quad
	\Rightarrow\quad {\bar X}'=\frac{\bphi'}{a^{2}}\left(\bphi''-\cH\bphi'\right)=2\bar X\left(\frac{\bphi''}{\bphi'}-\cH\right)
	\, ,
\ee
we can rewrite the background Eq. \rf{1.9} as
\be{1.11}
	\bphi''
	+\left(3c_s^2-1\right)\cH\bphi'
	+a^2\frac{
		\bar\varepsilon_{\phi}
	}{
		\bar\varepsilon_X
	}
	=0
	\, .
\ee
Alternatively, substituting the energy density $\bar \varepsilon$ by it definition \rf{1.5} in \rf{1.9} and multiplying the outcome by $a^2/\bphi'$, the evolution equation of $\bphi$ can be written in terms of the pressure $\bar P$ and its derivatives as follows
\be{1.12}
	\left(2\bar X{\bar P_{XX}}+{\bar P_X}\right)\bphi''
	-2\left(\bar X{\bar P_{XX}}-{\bar P_X}\right)\cH\bphi'
	+a^2\left(2\bar X{\bar P_{\phi X}}-{\bar P_{\phi}}\right)=0
	\, .
\ee
Before considering the perturbations of this background model, we will write some useful relations:
\be{1.13}
	\bphi''-\cH\bphi' = -3c_s^2\cH\bphi'
	-a^2\frac{
		\bar \varepsilon_{\phi}
	}{
		\bar \varepsilon_X
	}
	= -3\cH\frac{\bar P_X}{2\bar X\bar P_{XX}+\bar P_X}\bphi'
	-a^2\frac{
		2\bar X\bar P_{\phi  X}-\bar P_{\phi}
	}{
		2\bar X\bar P_{XX}+\bar P_X
	}
	\, ,
\ee
\ba{1.14}
{\bar P_X}&=&\frac{\bar\varepsilon+\bar P}{2\bar X}=\frac{a^{2}}{\left(\bphi'\right)^{2}}\left(\bar\varepsilon+\bar P\right)
	\, ,\\
	\label{1.15}
	\bar\varepsilon_X&=&\frac{\bar\varepsilon+\bar P}{2\bar Xc_s^2}=\frac{a^{2}}{\left(\bphi'\right)^{2}}\frac{\bar\varepsilon+\bar P}{c_s^2}
	\, .
\ea


\section{\label{sec:3}Scalar perturbations}

\setcounter{equation}{0}

Let us now consider a perturbed FLRW Universe with metric
\be{2.1}
\mathrm{d}s^{2}\approx a^{2}\left[(1+2\Phi)\mathrm{d}\eta^{2}-(1-2\Psi)\gamma_{\alpha\beta}\mathrm{d}x^{\alpha}\mathrm{d}x^{\beta}\right]\, .
\ee
The spatial sections have the metric $\gamma_{\alpha\beta}$ and their topology  can be flat ($\mathcal{K}=0$), closed or hyperbolic ($\mathcal{K}=\pm 1$). The perturbation of the $K$-essence scalar field energy-momentum tensor are
\ba{2.2}
\delta T_{0}^{0}
 & =&\frac{1}{a^{2}}\left(2\bar X{\bar P_{XX}}+{\bar P_{X}}\right)\left[\bphi'\pphi'-\left(\bphi'\right)^{2}\Phi\right]+\left(2\bar X{\bar P_{\phi X}}-{\bar P_{\phi}}\right)\pphi\, ,\\
 \label{2.3}
\delta T_{i}^{0} & =& =\frac{1}{a^{2}}{\bar P_{X}}\bphi'\dt_{i}\pphi\, ,\\
\label{2.4}
\delta T_{j}^{i} &=&-\delta_{j}^{i}\left[\frac{1}{a^{2}}{\bar P_{X}}\bphi'\pphi'-\frac{1}{a^{2}}{\bar P_{X}}\left(\bphi'\right)^{2}\Phi+{\bar P_{\phi}}\, \pphi\right]\, ,
\ea
where we split the scalar field into its background part $\bphi (\eta)$ and its fluctuations $\pphi (\eta,\bf r)$:
\be{2.5}
\phi = \bphi + \pphi\, .
\ee

Taking into account also the possible contributions of dust (which represents the inhomogeneities in the form of galaxies and groups of the galaxies) and radiation, Einstein equations for scalar perturbations are reduced to a system of three equations:
\ba{2.6}
&{}&\Delta\Phi-3\cH(\Phi'+\cH\Phi)+3\mathcal{K}\Phi =\frac{\kappa}{2}a^{2}\left(\delta\varepsilon_{\mathrm{dust}}+{\delta\varepsilon_{\mathrm{rad}}}\right)\nonumber \\
 &+&\frac{\kappa}{2}\left(2\bar X{\bar P_{XX}}+{\bar P_{X}}\right)\left[\bphi'\pphi'-\left(\bphi'\right)^{2}\Phi\right]+\frac{\kappa a^{2}}{2}\left(2\bar X{\bar P_{\phi X}}-{\bar P_{\phi}}\right)\pphi\, ,\\
&{}&\Phi'+\cH\Phi =\frac{\kappa}{2}{\bar P_{X}}\bphi'\pphi \label{2.7}\, ,\\
&{}&\frac{2}{a^{2}}\left[\Phi''+3\cH\Phi'+2\frac{a''}{a}\Phi-\cH^{2}\Phi-\mathcal{K}\Phi\right] \nn \\
& =&\label{2.8} \kappa\left[\delta P_{\mathrm{rad}}+\frac{1}{a^{2}}{\bar P_{X}}\bphi'\pphi'-\frac{1}{a^{2}}{\bar P_{X}}\left(\bphi'\right)^{2}\Phi+{\bar P_{\phi}}\, \pphi\right]\, ,
\ea
where $\kappa \equiv 8\pi G_N/c^4$ ($c$ is the speed of light and $G_N$ is the Newtonian gravitational constant). As we are in a regime  where the anisotropies can be neglected, we put $\Psi=-\Phi$. In Eq. \rf{2.6} $\Delta$ is the Laplace operator with respect to the metrics $\gamma_{\alpha\beta}$. For dust we have $\delta P_{\mathrm{dust}}=0$. As we mentioned above, we consider dust in the form of discrete distributed inhomogeneities (galaxies and groups of the galaxies) with the comoving real rest mass density:
\be{2.8a}
\rho = \frac{1}{\gamma}\sum_i m_i\delta({\bf r} - {\bf r}_i)\, ,
\ee
where $m_i$ and ${\bf r}_i$ are the mass and the comoving radius vector, respectively, of the $i$th inhomogeneity. For the fluctuation of the energy density of dustlike matter, we have \cite{Eingorn:2012jm}:
\be{2.8b}
\delta T^0_{0\; \mathrm{dust}} = \delta\varepsilon_{\mathrm{dust}}=\frac{\delta\rho c^2}{a^3}+\frac{3\bar\rho c^2\Phi}{a^3}\, .
\ee
where $\delta\rho$ is the difference between the real and the average rest mass densities (comoving):
\be{2.8c}
\delta\rho=\rho -\bar\rho\, .
\ee
In \rf{2.8b}, we keep only terms linear in $\Phi$  disregarding nonlinear expressions $\Phi^2/a^3, \Phi^3/a^3,$ etc (see footnote 2 in \cite{CPL}).
As we can see below (e.g. Eq. \rf{2.14}), $\Phi \sim 1/a$. Therefore, we drop all terms $O(1/a^n)$ with $n>4$. This is the accuracy of our approach. Radiation satisfies the well known EoS $\bar P_{\mathrm{rad}}=(1/3)\bar \varepsilon_{\mathrm{rad}}$. This results in the following behaviour $\bar\varepsilon_\mathrm{rad} \sim 1/a^4$.  Obviously, EoS for the fluctuations of radiation is $\delta p_{\mathrm{rad}}=\frac{1}{3}\delta\varepsilon_{\mathrm{rad}}$. Then, within the mechanical approach, we get (see, e.g., appendix in \cite{CPL}):  $\delta\varepsilon_{\mathrm{rad}}\sim 1/a^4$.

It is worth noting that according to the mechanical approach (see details in \cite{Eingorn:2012jm,EZcosm2}), we can drop the terms containing the peculiar velocities of the inhomogeneities and radiation as these are negligible when compared with their respective energy density and pressure fluctuations. However, when dealing  with a scalar field, such an approach is not evident since the quantity treated as the peculiar velocity of the scalar field is proportional to the scalar field perturbation $\varphi$. Therefore, in our analysis we propose the following strategy. 
As a first approximation, we set the scalar field perturbations \rf{2.7} to zero.
In a second and more accurate approach, we preserve the scalar field perturbation in \rf{2.7} since we keep it in Eqs. \rf{2.6} and \rf{2.8}. Then, a subsequent analysis of the equations must show whether or not we can equate to zero the right hand side (r.h.s.) of Eq. \rf{2.7}. In what follows, we shall demonstrate that for the coupled scalar field, r.h.s. of \rf{2.7} can indeed be set to zero in a consistent way within the mechanical approach as it usually happens for coupled fluids \cite{coupled}.

\subsection{Scalar field fluctuation $\pphi=0$\label{subsec3.1}}

Let us consider, first, the particular case $\pphi=0$. Even if the scalar field fluctuation is equal to zero, the fluctuations of its energy density and pressure do  not vanish due to the interaction between the $K$-essence scalar field background and the gravitational potential (see Eqs. \rf{2.2} and \rf{2.4}):
\ba{2.9}
\delta\varepsilon&=& -\frac{1}{a^{2}}\left(2\bar X{\bar P_{XX}}+{\bar P_{X}}\right)\left(\bphi'\right)^{2}\Phi\, ,\\
\label{2.10}
\delta P&=& -\frac{1}{a^{2}}{\bar P_{X}}\left(\bphi'\right)^{2}\Phi\, .
\ea
It is well known that energy densities and pressure are measurable quantities unlike fields. In the case $\varphi =0$, the Eqs. \rf{2.6}-\rf{2.8} read:
\ba{2.11}
&{}&\Delta\Phi-3\cH(\Phi'+\cH\Phi)+3\mathcal{K}\Phi =\frac{\kappa}{2}a^{2}\left(\delta\varepsilon_{\mathrm{dust}}+{\delta\varepsilon_{\mathrm{rad}}}\right)
-\frac{\kappa}{2}\left(2XP_{XX}+P_{X}\right)\left(\bphi'\right)^{2}\Phi\, ,\\
&{}&\Phi'+\cH\Phi  =0\ ,\label{2.12}\\
&{}&\frac{2}{a^{2}}\left[\Phi''+3\cH\Phi'+2\frac{a''}{a}\Phi-\cH^{2}\Phi-\mathcal{K}\Phi\right]  =\kappa\left[\frac{\delta\varepsilon_{\mathrm{rad}}}{3}-\frac{1}{a^{2}}P_{X}\left(\bphi'\right)^{2}\Phi\right]\, .\label{2.13}
\ea
From Eq. \rf{2.12} we obtain
\be{2.14}
\Phi(\eta,\textbf{r})=\frac{f(\textbf{r})}{c^{2}a(\eta)}\, .
\ee
where $f(\textbf{r})$ is a function of all spatial coordinates and we have introduced $c^2$ in the denominator for convenience.

After substituting  \rf{2.14} into \rf{2.11} and \rf{2.13}, we get
\ba{2.15}
&{}&\Delta f+3\mathcal{K}f  =\frac{\kappa c^{2}a^{3}}{2}\left(\delta\varepsilon_{\mathrm{dust}}+{\delta\varepsilon_{\mathrm{rad}}}\right)-
\frac{\kappa}{2}\left(2\bar X\bar P_{XX}+\bar P_{X}\right)\left(\bphi'\right)^{2}f\, ,\\
&{}&\left(\cH'-\cH^{2}-\mathcal{K}\right)\frac{f}{a^{3}}  =\frac{\kappa c^{2}}{2}\frac{\delta\varepsilon_{\mathrm{rad}}}{3}-\frac{\kappa}{2}\bar P_{X}\left(\bphi'\right)^{2}\frac{f}{a^{3}}\, .\label{2.16}
\ea
Using the background equation
\be{2.17}
\cH'-\cH^{2}-\mathcal{K}=-\frac{\kappa a^{2}}{2}
	\left(
		\bar\varepsilon_\textrm{dust}
		+\frac{4}{3}\bar\varepsilon_\textrm{rad}
		+\bar\varepsilon+\bar P
	\right)\, ,
\ee
in Eq. \rf{2.16}, we obtain
\be{2.18}
\left(\frac{4}{3}\bar\varepsilon_{\mathrm{rad}}+\bar\varepsilon_{\mathrm{dust}}\right)\frac{f}{a} =-\frac{c^{2}}{3}\delta\varepsilon_{\mathrm{rad}}\, ,
\ee
where we have used the relation:
\be{2.19}
\bar\varepsilon +\bar P =\left(\frac{\phi'_c}{a}\right)^2\bar P_X\, .
\ee
We now discard the $\bar\varepsilon_{\textrm{rad}}/a$ term as it decays with $(1/a)^5$ and is therefore outside of the accuracy of the method. Writing the background energy density
of dust as
\be{2.20}
\bar\varepsilon_{\mathrm{dust}}=\frac{\bar{\rho}c^{2}}{a^{3}}\, ,
\ee
we obtain from \rf{2.18}
\be{2.21}
\delta\varepsilon_{\mathrm{rad}}=-\frac{3\bar{\rho}}{a^{4}}f\,,
\ee
in full agreement with \cite{Eingorn:2012jm,EZcosm2}. This equation demonstrates that $\delta\varepsilon_{\mathrm{rad}}\sim 1/a^4$ as we mentioned already above.

Coming back to Eq. \rf{2.15} and substituting there Eq. \rf{2.8b}, we now find
\be{2.22}
\Delta f+3\mathcal{K}f  -\frac{\kappa c^{4}}{2}\delta\rho = -\frac{\kappa}{2}\left(2\bar X\bar P_{XX}+\bar P_{X}\right)\left(\bphi'\right)^{2}f\, .
\ee
In the previous equation, l.h.s. does not depend on time. In order for r.h.s. of this equation to be independent
of time we must impose
\be{2.23}
\left(2\bar X{\bar P_{XX}}+{\bar P_{X}}\right)\bar X=\frac{\beta^{2}}{2}\left(\frac{a_{0}}{a}\right)^{2}\, ,
\ee
where $\beta$ is a constant and where we have used \rf{1.10}. Taking into account this constraint and using the definition of the squared speed of sound (see Eq. \rf{1.7}), we also find that
\be{2.24}
c_{s}^{2}= \frac{2\bar X\bar P_X}{\beta^2}\left(\frac{a}{a_0}\right)^2\, .
\ee
Substituting Eq. \rf{2.23} in Eq. \rf{2.22}, we obtain
\be{2.25}
\Delta f-\lambda^{2}f=\frac{\kappa c^{4}}{2}\delta\rho\, ,
\ee
where $\lambda^{2}=-\frac{\kappa a_{0}^{2}}{2}\beta^{2}-3\mathcal{K}$ and $\delta\rho$ is defined in Eqs. \rf{2.8a} and \rf{2.8c}. This equation can be solved for any spatial topology \cite{Burgazli:2013qy}. For example, in the case of spatially flat ($\mathcal{K}=0$) and hyperbolic ($\mathcal{K}=-1$) geometries
we get, respectively:
\ba{2.26}
f&=&-G_{N}\underset{i}{\sum}\frac{m_{i}}{\vert\mathbf{r}-\mathbf{r}_{i}\vert}\exp\left(-\lambda\vert\mathbf{r}-\mathbf{r}_{i}\vert\right)+\frac{4\pi G_{N}\bar{\rho}}{\lambda^{2}}\, ,\\
\label{2.27}
f&=&-G_N\sum_i m_i\frac{\exp(-l_i\sqrt{\lambda^2+1}\, )}{\sinh l_i}+\frac{4\pi G_N\overline\rho}{\lambda^2}\, ,
\ea
where $l_i$ denotes the geodesic distance between the $i$-th mass $m_i$ and the point of observation. To obtain these physically reasonable  solutions (i.e. solutions which have the correct Newtonian limit near inhomogeneities and converge
at spatial infinity) \rf{2.26}, we impose that $\lambda^2 >0$. In the case of spatially flat topology, this means that $\beta^2<0$. However, for a hyperbolic space, $\lambda^2$ can acquire positive values if $\beta^2$ is positive.

Eqs. \rf{2.9} and \rf{2.10} demonstrate that $\delta\varepsilon$ and $\delta P$ are proportional to the gravitational potential $\Phi =f/(c^2 a)$. Therefore, as it follows from Eqs. \rf{2.26} and \rf{2.27}, these fluctuations are concentrated around the inhomogeneities screening them. Hence, the considered $K$-essence scalar field is really coupled to the galaxies and the group of galaxies. Moreover, as it follows from \rf{2.21}, the fluctuations of radiation are also coupled to the inhomogeneities.

One further very important feature of the gravitational potentials \rf{2.26} and \rf{2.27}:
as it was shown in \cite{Burgazli:2013qy}, these solutions satisfy the important condition that the total gravitational potential averaged over the whole Universe is equal to zero: $\overline f =0 \Rightarrow \overline \Phi =0$. This implies another physically reasonable condition: $\overline{\delta\varepsilon}=\overline{\delta P}=0$ (see Eqs.~\rf{2.9} and \rf{2.10}). It is obvious that the averaged value of the fluctuations should be equal of zero.

In order to solve Eqs. \rf{1.12} and \rf{2.23}, we need to define the explicit form of the function $P(X,\phi)$. For example, in the case of the canonical minimally coupled scalar field:
\be{2.28}
P\left(X,\phi\right)=X-V\left(\phi\right) \quad \Rightarrow \quad {P_{X}}=1\,,\quad{P_{\phi}}=-\frac{\dt V}{\dt\phi}\,,\quad{P_{XX}}={P_{\phi X}}=0\, ,
\ee
we reproduce the results of the paper \cite{Burgazli:2015mzm}:
\be{2.29}
\phi_c = \beta a_0\eta +\gamma\, ,\quad V(a)=\beta^2\left(\frac{a_0}{a}\right)^2+V_{\infty}\, , \quad \gamma, V_{\infty} = const\, ,
\ee
where in \cite{Burgazli:2015mzm}, we set $V_{\infty}$ to zero (see lhs column of page 5 in \cite{Burgazli:2015mzm}).
Below, we shall consider some more general and interesting examples.

\subsection{General scalar field fluctuation $\pphi$}

Now,  r.h.s. of Eq. \rf{2.7} is not equal to zero. Differentiating this equation with respect to the conformal time, we obtain
\be{2.30}
\Phi''+\cH\Phi'+\cH'\Phi  =\frac{\kappa}{2}{\bar P_{X}}\left[\bphi'\pphi'-2\cH\bphi'\pphi\right]+\frac{\kappa a^{2}}{2}{\bar P_{\phi}}\pphi\, ,
\ee
where we have also used the relation \rf{1.13}. Substituting Eqs. \rf{2.7} and \rf{2.30} in \rf{2.8} we get
\be{2.31}
\frac{2}{a^{2}}\left(\cH'-\cH^{2}-\mathcal{K}\right)\Phi  =\kappa\left[\delta p_{\mathrm{rad}}-\frac{1}{a^{2}}{\bar P_{X}}\left(\bphi'\right)^{2}\Phi\right]\, .
\ee
This equation looks similar to Eq. \rf{2.16}. However, in this case the dependence of $\Phi$ on the scale factor is not specified in \rf{2.31}.
With the help of Eqs. \rf{2.17} and \rf{1.14}, we find
\be{2.32}
-\left(\frac43\bar \varepsilon_{\textrm{rad}}+\bar\varepsilon_{\textrm{dust}}\right)\Phi  =\frac13 \delta \varepsilon_{\textrm{rad}}\, .
\ee
Since $\bar \varepsilon_{\textrm{rad}}\sim 1/a^4$, then for large scale factors, $a$, we can drop this term  as compared to $\bar \varepsilon_{\textrm{dust}}\sim 1/a^3$. Moreover, as we shall see later (c.f. the last paragraph of the current subsection), the product $\frac43\bar \varepsilon_{\textrm{rad}} \Phi\sim O(1/a^5)$ for the physically solvable case. Therefore, $\frac43\bar \varepsilon_{\textrm{rad}} \Phi$ can be neglected as it is out of the accuracy of the mechanical approach. Therefore, our consideration is self-consistent and we arrive at
\be{2.33}
\delta\varepsilon_{\mathrm{rad}}=-3\bar\varepsilon_{\textrm{dust}}\Phi\, .
\ee
This is a physically reasonable expression for the radiation fluctuations coupled to galaxies and cluster of galaxies (see, e.g., Eq. (4.19) in \cite{EZcosm2}).

Let us now go back to Eq. \rf{2.6}. Taking into account Eqs. \rf{2.7}, \rf{2.8b}, and \rf{2.33}, we get
\be{2.34}
\Delta\Phi+3\mathcal{K}\Phi-\frac{\kappa c^{2}}{2}\frac{\delta\rho}{a} =\frac{\kappa}{2}\left(2\bar X\bar P_{XX}+{\bar P_{X}}\right)\left[\bphi'\pphi'-\left(\bphi'\right)^{2}\Phi-\left(\bphi''-\cH\bphi'\right)\pphi\right]\, ,
\ee
which is complemented by Eq. \rf{2.7}.

We next use the ansatz
\be{2.35}
\Phi(\eta,{\bf{x}})\equiv \frac{\Omega(\eta,{\bf{x}})}{a}\, ,\quad
\pphi(\eta,{\bf{x}})\equiv \frac{\bphi'}{a}\chi(\eta,{\bf{x}})\, .
\ee
From this ansatz, we get:
\ba{2.36}
\Omega' & =& a\left(\Phi'+\cH\Phi\right)\, ,\\
\chi' & =&\frac{a}{(\bphi')^{2}}\left[\bphi'\pphi'-\left(\bphi''-\cH\bphi'\right)\pphi\right]\, .\label{2.37}
\ea
Then, Eqs. \rf{2.34} and \rf{2.7} read, respectively:
\ba{2.38}
&{}&\Delta\Omega+3\mathcal{K}\Omega-\frac{\kappa c^{2}}{2}\delta\rho =\frac{\kappa}{2}\left(2\bar X\bar P_{XX}+{\bar P_{X}}\right)\left(\bphi'\right)^{2}\left[\chi'-\Omega\right]\, ,\\
&{}&\Omega'  =\frac{\kappa}{2}{\bar P_{X}}(\bphi')^{2}\chi\, .\label{2.39}
\ea
Differentiating the second equation, we find
\be{2.40}
\chi' = \frac{2}{\kappa}\frac{1}{\bar P_{X}(\bphi')^{2}}\left[\Omega''-\left(\frac{2\bar X\bar P_{XX}+\bar P_{X}}{\bar P_{X}}\left(\frac{\bphi''}{\bphi'}-\cH\right)+\frac{\bphi''}{\bphi'}+\cH+\frac{\bar P_{\phi X}}{\bar P_{X}}\bphi'\right)\Omega'\right]\, .
\ee
Substituting this expression in Eq. \rf{2.38}, we obtain a second order
linear partial differential equation for $\Omega$ with the source term
$(\kappa c^{2}/2)\delta\rho$:
\be{2.41}
\Delta\Omega+3\mathcal{K}\Omega-\frac{\kappa c^{2}}{2}\delta\rho =\frac{1}{c_{s}^{2}}\left[\Omega''-\left(\frac{c_{s}^{2}+1}{c_{s}^{2}}\frac{\bphi''}{\bphi'}+\frac{c_{s}^{2}-1}{c_{s}^{2}}\cH+\frac{\bar P_{\phi X}}{\bar P_{X}}\bphi'\right)\Omega'-\frac{\kappa}{2}{\bar P_{X}}(\bphi')^{2}\Omega\right]\, .
\ee
Here we have used the definition of $c_{s}^{2}$ given in Eq. \rf{1.7}.

Let us now analyse the structure of this equation. If
$c_{s}^{2}>0$ then the previous equation corresponds to an inhomogeneous
hyperbolic partial differential equation. In fact, by multiplying it
by $c_{s}^{2}$ and rearranging the terms we obtain a general wave
equation
\be{2.42}
{c_{w}^{2}(\eta)}\Delta\Omega(\eta,\vec{x})-\Omega''(\eta,\vec{x})-\gamma(\eta)\Omega'(\eta,\vec{x})-k(\eta)\Omega(\eta,\vec{x})=
F(\vec{x})\, ,
\ee
where we identify the quantities:
\begin{enumerate}
\item $c_{w}={c_{s}}$ is the propagation speed of the waves;
\item $\gamma=-\left(\frac{c_{s}^{2}+1}{c_{s}^{2}}\frac{\bphi''}{\bphi'}+\frac{c_{s}^{2}-1}{c_{s}^{2}}\cH+\frac{P_{\phi X}}{P_{X}}\bphi'\right)$
is the damping factor;
\item $k=-\left[3\mathcal{K}+\frac{\kappa}{2}{P_{X}}(\bphi')^{2}\right]$
is the restoration factor;
\item $F=\frac{\kappa c^{2}}{2}\delta\rho$ is the source term.
\end{enumerate}
The solution to Eq. \rf{2.42} can be written as
\be{2.43}
\Omega=\Omega_{\textrm{wave}}+\Omega_{\delta\rho}\, ,
\ee
where $\Omega_{\textrm{wave}}$ corresponds to the travelling wave
that solves the homogeneous equation. We should notice that this solution (regardless of the sign of $c_s^2$) is
``blind'' to the effects of the source term. Therefore, this solution lacks physical  interest because we are studying fluctuations arising precisely from the inhomogeneities. Second, it is obvious that the travelling waves cannot be coupled to the inhomogeneities. Therefore, we should set $\Omega_{\textrm{wave}}\equiv 0$. Obviously, a particular solution $\Omega_{\delta\rho}$ of the inhomogeneous equation encodes all the information regarding the source term. A physically reasonable solution corresponds to the assumptions $\Omega_{\delta\rho}\equiv f({\bf{x}})$ and
\be{2.44}
	c_s^2 = \frac{\bar P_X}{2\bar X \bar P_{XX}+\bar P_X} = \left(\frac{\bphi'}{a_0\beta}\right)^2\bar P_X\, , \quad \beta=\mathrm{const}
	\, .
\ee
The later one is similar to the physical case described by \rf{2.23}. Such a choice provides, e.g., a correct transition to the astrophysical approach where the gravitational potential has the right Newtonian form \cite{Burgazli:2015mzm}. Under these assumptions, the function  $f({\bf{x}})$ satisfies Eq. \rf{2.25}. Additionally, since the function $\Omega$ does not depend on time, we get that $\chi =0 \Rightarrow \varphi =0$ (see Eq. \rf{2.39}) and all the results obtained on the previous subsection 3.1 remain valid.

\

\section{\label{sec:4}Particular examples}

\setcounter{equation}{0}

\

In order to retrieve more information for $K$-essence models within the mechanical approach, we need to specify the algebraic form of the $K$-essence Lagrangian, $P(X,\phi)$. Our analysis will be also carried within the framework of subsection \ref{subsec3.1}. In addition,  in this section, we consider models where the function $P$ has the general form
\be{3.1a}
P=K\left(X\right)-V\left(\phi\right)\, .
\ee
In this kind of models, the equations of the $K$-essence simplify considerably since any term with crossed derivatives $P_{\phi X}$ vanishes automatically. Furthermore the constraint \rf{2.23} can be used to determine the dependence of the kinetic term on the size of the Universe, i.e. $X(a)$. We can also find a relation $V(a)$ from Eq. \rf{1.11} which can be re-written in the form:
\be{3.1}
V_a  \equiv \frac{\mathrm{d}V}{\mathrm{d}a}
	=-\frac{\beta^2}{2}\left(\frac{a_0}{a}\right)^2\left[\frac{6c_s^2}{a}+\frac{1}{\bar X}\frac{\mathrm{d} \bar X}{\mathrm{d} a}\right]\, .
\ee
This relation has been obtained by taking into account that $\bar P_\phi=-V_\phi=-(a\cH/\phi_c')(\mathrm{d} V_a/\mathrm{d} a)$, using Eq. \rf{2.23} and the auxiliary  equations:
$$
\bphi^{\prime}=a\sqrt{2\bar X}\, , \quad \bphi^{\prime\prime}=\mathcal{H}\bphi^{\prime}\left(1+\frac{a^{3}}{\bphi^{\prime2}}\frac{d \bar X}{d a}\right)\, .
$$

In what follows, we study relevant $K$-essence models by analysing the solutions of Eqs.~\rf{2.23} and ~\rf{3.1}.

\subsection{Purely kinetic $K$-essence (PKK) models}

\subsubsection{Generic case}

Purely kinetic $K$-essence (PKK) models can in general be described
by a Lagrangian with the form
\be{3.2}
P\left(X,\phi\right)=P\left(X\right)\, .
\ee
These type of models were considered as a possible resolution
of the late-time cosmic acceleration problem (see for example Refs.~\cite{Chiba:1999ka,ArmendarizPicon:2000dh,Chimento:2003zf,Chimento:2003ta,dePutter:2007ny,
Scherrer:2004au,BouhmadiLopez:2010vi}). 
As we shall see below, the PKK models with the coupled $K$-essence scalar field can really provide the late-time acceleration of the Universe; i.e. a late-time acceleration of the Universe compatible with the scalar perturbations as dictated by the mechanical approach.

The equation of motion \rf{1.9} for a PKK field takes the form
\be{3.3}
\left(\bar P_{X}+2\bar X\bar P_{XX}\right)\dot{\bar X}+6H\bar X\bar P_{X}=0\, ,\
\ee
where the dot denotes the derivative with respect to the cosmic (synchronous)  time ($t$) and $H$ is the Hubble parameter; i.e.  $H=\dot{a}/a$.
Changing the independent variable from the cosmic time $t$ to the
scale factor $a$, Eq. \rf{3.3} can be re-written as \cite{Scherrer:2004au}
\be{3.4}
\left(\bar P_{X}+2\bar X\bar P_{XX}\right)a\frac{\mathrm{d}\bar X}{da}+6\bar X\bar P_{X}=0\, .
\ee
For $\bar P_X\neq$ const, the solution of this equation is
\be{3.5}
\bar X\bar P_{X}^{2}=k\left(\frac{a}{a_{0}}\right)^{-6}\Rightarrow \bar P_{X}=\sqrt{\frac{k}{\bar X}}\left(\frac{a}{a_{0}}\right)^{-3}\, ,
\ee
where $k$ is an integration constant. Therefore, for the speed of sound squared \rf{2.24} we obtain
\be{3.6}
	c_s^2 = \frac{2}{\beta^2}\sqrt{k \bar X}\left(\frac{a_0}{a}\right) = (c_s^2)_0\; \sqrt{\frac{\bar X}{\bar X_0}}\, \left(\frac{a_0}{a}\right)\, .
\ee
The present time values $(c_s^2)_0$ and $\bar X_0$ are expressed via the free parameters $k$ and $\beta^2$ as
\be{3.7}
\frac{\sqrt{k}}{\beta^2}= \frac{1}{2}\frac{\left(c_{s}^{2}\right)_0}{\sqrt{\bar X_0}}\, .
\ee

Substituting Eqs. \rf{2.23} and \rf{3.5} in Eq. \rf{3.4}, we obtain
\be{3.8}
\frac{\mathrm{d}\bar X}{\bar X\sqrt{\bar X}}=-\frac{12\sqrt{k}a_{0}}{\beta^{2}} \frac{\mathrm{d}a}{a^2}\, .
\ee
The solution of this equation reads
\be{3.9}	
\bar X = \bar X_0 \left[1+ 3\left(c_{s}^{2}\right)_0\left(1-\frac{a_0}{a}\right)\right]^{-2}\, .
\ee
If $(c_s^2)_0=-1/3$ the previous expression reduces to $X=X_0(a/a_0)^2$, and therefore the kinetic term $X$ blows up in the distant future as $a\rightarrow\infty$. In this case, the Universe is asymptotically de Sitter as can be easily deduced from the expressions (\ref{3.15}) and (\ref{3.16}). For $(c_s^2)_0\neq-1/3$ we find that in the limit of very large scale factors the kinetic variable $X$ does not decay to zero and instead goes to a constant value $X_\infty= X_0[1 + 3(c_{s}^{2})_0]^{-2}$. However, for $(c_s^2)_0<-1/3$ we find that $X$ diverges at a finite value of $a>a_0$ while for $ (c_s^2)_0>0$ we find that $X$ diverges at a finite value $a<a_0$. It can be shown that in this case, the Universe would face a big freeze singularity \cite{BouhmadiLopez:2006fu,BouhmadiLopez:2007qb}; i.e. a dark energy singularity at a finite scale factor. Since the methods developed so far in the mechanical approach should be applicable for big values of the scale factor $a>>a_0$, we, therefore, restrict our analysis to $(c_s^2)_0\geq -1/3$.

With the help of Eq. \rf{3.9}, the formula \rf{3.6} for the  speed of sound squared reads
\be{3.10}
	c_s^2
	=  \frac{\left(c_s^2\right)_0\left(\frac{a_0}{a}\right)}{1+ 3\left(c_{s}^{2}\right)_0\left(1-\frac{a_0}{a}\right)}
	\, .
\ee
In the case of $(c_s^2)_0=-1/3$, the right hand side of this equation reduces to a constant with $c_s^2=(c_s^2)_0=-1/3$. Furthermore, similar to what was found for $X$, for $(c_s^2)_0<-1/3$, the  speed of sound squared has a divergence in the future, while for $(c_s^2)_0>0$ the  speed of sound squared has a divergence in the past. For $-1/3<(c_s^2)_0<0$ the  speed of sound squared is initially $-1/3$ at $a=0$ and raises monotonically to $-0$ in the distant future as $a\rightarrow \infty$.

Let us now invert the relation \rf{3.9} to obtain the ratio $a_0/a$ as a function of $X$:
\be{3.11}
	\frac{a_0}{a} =1 + \frac{1}{3\left(c_{s}^{2}\right)_0}\left[1  - \left(\frac{\bar X}{\bar X_0}\right)^{-1/2}\right]
	\, .
\ee
By substituting this equation into  Eq. \rf{3.5}, we get the differential equation
\be{3.12}
	P_{X}=\sqrt{\frac{k}{\bar X}}
	\left\{1 + \frac{1}{3\left(c_{s}^{2}\right)_0}\left[1  - \left(\frac{\bar X}{\bar X_0}\right)^{-1/2}\right]\right\}^{3}\, ,
\ee
which has the solution:
\ba{3.13}
	\bar P  &=& \bar P_0 + \frac{\beta^2}{6}\left(\frac{1}{3\left(c_{s}^{2}\right)_0}\right)^2
	\left\{
		2\left[1 + 3\left(c_{s}^{2}\right)_0\right]^3
		\left[\left(\frac{\bar X}{\bar X_0}\right)^{1/2}-1\right]
		-3 \left[1 + 3\left(c_{s}^{2}\right)_0\right]^2
		\log\left(\frac{\bar X}{\bar X_0}\right)
	\right. \nn\\
	&-&\left.
		6 \left[1 + 3\left(c_{s}^{2}\right)_0\right]
		\left[\left(\frac{\bar X}{\bar X_0}\right)^{-1/2}-1\right]
		+\left[\left(\frac{\bar X}{\bar X_0}\right)^{-1}-1\right]\right\}
	\, .
\ea
Using the definition of the energy density of the $K$-essence \rf{1.5}, we obtain
\ba{3.14}
	\bar \varepsilon &=& 2\bar X\bar P_X-\bar P
	=\varepsilon_0 + \frac{\beta^2}{6}\left(\frac{1}{3\left(c_{s}^{2}\right)_0}\right)^2
	\left\{
		3 \left[1 + 3\left(c_{s}^{2}\right)_0\right]^2
		\log\left(\frac{\bar X}{\bar X_0}\right)
	\right. \nn\\
	&+&\left.
		12 \left[1 + 3\left(c_{s}^{2}\right)_0\right]
		\left[\left(\frac{\bar X}{\bar X_0}\right)^{-1/2}-1\right]
		-3\left[\left(\frac{\bar X}{\bar X_0}\right)^{-1}-1\right]
	\right\}
	\, ,
\ea
where we have introduce the value of the energy density of the $K$-essence at the present time, defined as $\bar \varepsilon_0 = -\bar P_0 + \beta^2(c_s^2 )_0$.

In the particular case of $(c_s^2)_0=-1/3$, we obtain
\be{3.15}
	\bar P = \bar P_0 + \frac{\beta^2}{6}\left[\left(\frac{\bar X}{\bar X_0}\right)^{-1}-1\right]
	= \bar P_0 + \frac{\beta^2}{6}\left[\left(\frac{a_0}{a}\right)^{2}-1\right]
	\, ,
\ee
which has the limiting value $\bar P_\infty=\bar P_0-\frac{\beta^2}{6}$, and
\be{3.16}
	\bar \varepsilon = - \bar P_0 + \frac{\beta^2}{6}\left[1 - 3\left(\frac{\bar X}{\bar X_0}\right)^{-1}\right]
	= - \bar P_0 + \frac{\beta^2}{6}\left[1 - 3\left(\frac{a_0}{a}\right)^{2}\right]
	\,,
\ee
with the limiting value $\bar \varepsilon_\infty = -\bar P_\infty$. If $\bar P_\infty\neq0$, the $K$-essence behaves (at the background level) as a two-components perfect fluid: a cosmological constant and a network of frustrated cosmic strings (c.f. for example \cite{FNCS}). In fact, the effective EoS reads
\be{3.16a}
	\bar P = \dfrac{2}{3}\bar{P}_0 - \dfrac{1}{9}\beta^2 - \dfrac{1}{3}\bar{\varepsilon}
	\,.
\ee
Therefore, in this particular case the coupled $K$-essence scalar field can describe the late-time acceleration of the Universe.  If $\bar P_\infty=0$, the $K$-essence perfectly mimics a network of frustrated cosmic strings with $w=c_s^2=-1/3$.

Now, we analyse the late-time behaviour of the $K$-essence for a general value of $c_s^2\neq-1/3$. To perform it, we  can rewrite Eqs. \rf{3.13} and \rf{3.14} as follows:
\ba{3.17}
\bar P
	&=& \bar P_\infty + \frac{\beta^2}{6}\left[1 - \left(\frac{\bar X_\infty}{\bar X_0}\right)^{1/2}\right]^{-2}
	\left\{
		2\left[\left(\frac{\bar X}{\bar X_\infty}\right)^{1/2} - 1\right]\right.\nn\\
&-&\left. 3\log\left(\frac{\bar X}{\bar X_\infty}\right)
		-6 \left[\left(\frac{\bar X}{\bar X_\infty}\right)^{-1/2} - 1\right]
		+\left[\left(\frac{\bar X}{\bar X_\infty}\right)^{-1} -  1\right]
	\right\}
	\, ,
\ea
and
\be{3.18}
\bar\varepsilon
=\bar \varepsilon_\infty + \frac{\beta^2}{6}\left[1 - \left(\frac{\bar X_\infty}{\bar X_0}\right)^{1/2}\right]^{-2}
	\left\{
		3 \log\left(\frac{\bar X}{\bar X_\infty}\right)
		+12
		\left[\left(\frac{\bar X}{\bar X_\infty}\right)^{-1/2} -1\right]
		-3\left[\left(\frac{\bar X}{\bar X_\infty}\right)^{-1} - 1\right]
	\right\}
	\, .
\ee
Here,  we have introduced the limiting values of $X$ for large scale factors for which  the energy density and the pressure can be expressed as:
\ba{3.19}
\bar X_\infty&=&\bar X_0[1+3(c_s^2)_0]^{-2}\, ,\\
\bar P_\infty  &=& \bar P_0 + \frac{\beta^2}{6}\left[1 - \left(\frac{\bar X_\infty}{\bar X_0}\right)^{1/2}\right]^{-2}
	\left\{
		2\left[1 - \left(\frac{\bar X_\infty}{\bar X_0}\right)^{-1/2}\right]\right.\nn\\
&-&\left.3\log\left(\frac{\bar X_\infty}{\bar X_0}\right)
		-6 \left[1- \left(\frac{\bar X_\infty}{\bar X_0}\right)^{1/2}\right]
		+ \left[1 - \left(\frac{\bar X_\infty}{\bar X_0}\right)\right]
	\right\}\label{3.20}\, ,\\
\bar\varepsilon _\infty
	&=&\bar \varepsilon_0 + \frac{\beta^2}{6}\left[1 - \left(\frac{\bar X_\infty}{\bar X_0}\right)^{1/2}\right]^{-2}
	\left\{
		3 \log\left(\frac{\bar X_\infty}{\bar X_0}\right)\right. \nn\\
&+&\left.12\left[1 - \left(\frac{\bar X_\infty}{\bar X_0}\right)^{1/2}\right]
		-3\left[1 - \left(\frac{\bar X_\infty}{\bar X_0}\right)\right]
	\right\}=-\bar P_\infty\label{3.21}\, .
\ea
From these expressions, we find that $P_\infty+\varepsilon_\infty=0$. Thus, if $P_\infty=-\varepsilon_\infty\neq0$, the $K$-essence field behaves as a cosmological constant at late time. Hence, in the case $c_s^2\neq-1/3$, we again arrive to the conclusion that the coupled $K$-essence can provide the current cosmic acceleration. To describe the considered model in more detail, we examine the next leading order terms in the expansion $1/a$. To do that, we rewrite Eq. \rf{3.9} as
\be{3.22}
	\left(\frac{\bar X}{\bar X_\infty}\right)  = \left(1 + \left[\left(\frac{\bar X_\infty}{\bar X_0}\right)^{1/2} - 1\right]\left(\frac{a_0}{a}\right)\right)^{-2}
	\, ,
\ee
substitute this expressions in Eqs. \rf{3.17} and \rf{3.18} and expand them  around $a\rightarrow\infty$:
\ba{3.23}
\bar P  &=& \bar P_\infty + \frac{\beta^2}{6}\left[1 - \left(\frac{\bar X_\infty}{\bar X_0}\right)^{1/2}\right]^{-2}
	\left\{
		\frac{1}{3}A_{r}\left(\frac{a_0}{a}\right)^4
		 + O\left(\frac{1}{a^5}\right)
	\right\}\, ,\\
\bar\varepsilon
	&=&\bar \varepsilon_\infty + \frac{\beta^2}{6}\left[1 - \left(\frac{\bar X_\infty}{\bar X_0}\right)^{1/2}\right]^{-2}
	\left\{
		A_{d}\left(\frac{a_0}{a}\right)^3
		+A_{r}\left(\frac{a_0}{a}\right)^4
		 + O\left(\frac{1}{a^5}\right)
	\right\}\label{3.24}\, ,
\ea
where
\be{3.25}
	A_d \equiv  2\left[1 - \left(\frac{\bar X_\infty}{\bar X_0}\right)^{1/2}\right]^3\, ,\quad
	A_r \equiv  \frac{3}{2}\left[1 - \left(\frac{\bar X_\infty}{\bar X_0}\right)^{1/2}\right]^4\, .
\ee
We thus find that if $c_s^2 > -1/3$ the PKK model compatible with the constraint of the mechanical approach \rf{2.23} behaves at late-time like a fluid with three components: radiation, dust, and a  cosmological constant.
Since Eqs. \rf{3.23} and \rf{3.24} were derived with the help of Eqs. \rf{2.23} and \rf{2.24}, the cosmological constant is a consequence of the condition of the coupling of the $K$-essence scalar field to the "discrete" and inhomogeneous structures in the Universe; i.e. galaxies. We did not put such a constant by hand in the action. It is clear that such a model can provide the late-time cosmic acceleration of the Universe in the absence of a scalar field potential term.
Obviously, radiation and dust are dark (in the sense that they mimic the ordinary dust and radiation). The ratios between each of the components depend on the parameter $\beta^2$ and on the value of the speed of sound squared of the $K$-essence at present $(c_s^2)_0$. As our results above show, in order for the model to be free of divergences in the evolution of the kinetic variable $X$, we must constraint the model to satisfy $-1/3\leq(c_s^2)_0\leq0$. This means that for the coupled $K$-essence scalar field (in the considered PKK models) the speed of sound is imaginary. Usually, this leads to ghost instabilities \cite{Amendola:2004qb,Piazza:2004df,delaMacorra:2002du,Erickson:2001bq}. However, as it was shown in \cite{BS,BCCM}, perfect fluids with negative speed of sound squared could be stable if they are sufficiently rigid. This is our case because, due to the concentration of fluctuations of the coupled perfect fluids around the inhomogeneities (e.g. galaxies), they have velocities of the order of these inhomogeneities, i.e. very non-relativistic. Obviously, the physical speed of sound in this case is close to zero. As noted on the paper \cite{Conversi}, for the "solid" dark energy, a zero speed of sound is preferable.

\

\subsubsection{Generalized Chaplygin gas model as a PKK model}

\

The pure kinetic $K$-essence models, as the name indicates, are those
$K$-essence models whose Lagrangian depends only on the kinetic variable
$X$ and not directly on the field itself (see Eq. \rf{3.2}).
These kinds of models have the particularity that the field $\phi$
is absent from the evolution equations at the background level. Only
$\phi'$ and derivatives of higher orders appear in these equations.
In practice, this eliminates one degree of freedom from the system
and, as long as $\bar\varepsilon(\bar X)$ is an invertible function, we find
that a PKK model is equivalent to a perfect fluid with barotropic
EoS $\bar P=-\bar\varepsilon+2\bar X(\bar\varepsilon)\bar P_{X}(\bar \varepsilon)$ (see Eq. \rf{1.5}).
Outside the scope of the $K$-essence models various examples of perfect
fluids with barotropic equations of state have previously been studied
within the Mechanical Approach as candidates for Dark Energy \cite{Burgazli:2013qy,Bouhmadi-Lopez:2015oxa,Burgazli:2015mzm,CPL,ENZ1,Laslo2}.

The fact that the pressure does not depend on the value of the
field $\phi$ means that in a PKK model the last term of equation
\rf{1.12} vanishes and we can solve this equation 
to obtain the relation \rf{3.5}, where $k=\sqrt{\bar X_{0}}\bar P^2_{X}(\bar X_{0})$.
Once the shape of $\bar P(\bar X)$ is specified we can invert Eq. \rf{3.5}
in order to obtain a unique solution $\bar X=\bar X(a)$ with $\bar X(a_{0})=\bar X_{0}$.
This in turn allows us to write the pressure $\bar P$, as well as its
derivatives $\bar P_{X}$ and $\bar P_{XX}$, as a function of the scale factor%
\footnote{Notice that this possibility of writing explicitly the dependence
on the scale factor of all the relevant quantities of the $K$-essence
field is a striking difference between the PKK and the general $K$-essence
models; in the latter usually such solutions are only found by
means of approximations in specific regimes, e.g., the potential dominated
slow-roll of a canonical scalar-field.%
} and then expand the right hand side of Eq. \rf{2.22} in
powers of $(1/a)$, while keeping only those terms that comply with
the accuracy of the method. Hence, in a PKK scenario we dot need to
impose additionally the constraint \rf{2.23} in order to analyse the
compatibility of the equations of linear scalar perturbations of the
$K$-essence within the Mechanical Approach.

As an example we take the case of the Generalised Chaplygin Gas  (GCG)
\cite{Kamenshchik:2001cp,Bilic:2001cg,Bento:2002ps}, which has previously been studied in the Mechanical
Approach in \cite{Bouhmadi-Lopez:2015oxa}. As derived in \cite{Bento:2002ps}
the GCG can be interpreted as a $K$-essence field with the Lagrangian (see \cite{BouhmadiLopez:2004me} for the complementary phantom case)
\be{3.26}
\mathcal{L}_{GCG}=-A^{\frac{1}{1+\alpha}}\left[1-X^{\frac{1+\alpha}{2\alpha}}\right]^{\frac{\alpha}{1+\alpha}}\, .
\ee
It can now be recognized that the GCG Lagrangian \rf{3.26}
falls within the category of PKK models. Solving Eq. \rf{3.5}
with $P(X)=\mathcal{L}_{GCG}$, we obtain
\be{3.27}
\bar X=\left[1+\left(\bar X_{0}^{-\frac{1+\alpha}{2\alpha}}-1\right)\left(\frac{a}{a_{0}}\right)^{3(1+\alpha)}\right]^{-\frac{2\alpha}{1+\alpha}}\, ,
\ee
while expanding Eq. \rf{2.22} leads to
\ba{3.28}
&{}&\Delta f+3\mathcal{K}f  -\frac{\kappa c^{4}}{2}\delta\rho = -\kappa a^{2}\left(2\bar X^{2}\bar P_{XX}+\bar X\bar P_{X}\right)f\nn\\
& =& -\frac{\kappa a^{2}}{2}\frac{A^{\frac{1}{1+\alpha}}}{\alpha}1+\frac{1}{\bar X_{0}^{-\frac{1+\alpha}{2\alpha}}
-1}\left(\frac{a}{a_{0}}\right)^{-3(1+\alpha)}
\left[1+\frac{1}{\bar X_{0}^{-\frac{1+\alpha}{2\alpha}}-1}\left(\frac{a}{a_{0}}\right)^{-3(1+\alpha)}\right]^{\frac{1}{1+\alpha}}f\, .\nn\\
&{}&
\ea
This expression is equivalent to Eqs. (3.2) and (3.4) of \cite{Bouhmadi-Lopez:2015oxa} with $\beta=0$ and $\delta\varepsilon_{rad2}=0$
in that work. Therefore, we can simply apply the results found there.

\

\subsection{Scalar field with constant sound speed}

\

In this subsection, we study  $K$-essence models with constant sound
speed. Such  models are defined by the Lagrangian \cite{Sergijenko:2014pwa}
\be{3.29}
P\left(X,\phi\right)=U\left(\phi\right)X^{\gamma}-V\left(\phi\right)\, ,
\ee
where $\gamma=(1+c_{s}^{2})/2c_{s}^{2}=\mathrm{const}\neq 1/2$, $U\left(\phi\right)$
and $V\left(\phi\right)$ are arbitrary functions of $\phi$. For
$U\left(\phi\right)=1$ and $c_{s}^{2}=1$, we recover the standard
canonical scalar field. In Ref. \cite{Sergijenko:2014pwa}, it is
concluded (on the base of CMB anisotropies and BAO data) that the sound speed is unconstrained in the range $\left[0\,,\,1\right]$
for the model defined by Eq. \rf{3.29}.

In general, models of the form \rf{3.29} do not belong to the class described by the action \rf{3.1a}. However, for simplicity, we consider the particular case
$U\left(\phi\right)=1$. Therefore, such a model is within the scope of the present section.  With the help of the mechanical approach, we want to obtain now
the potential $V$, pressure $\bar P$ and the energy density $\bar\varepsilon$ as functions of the scale factor $a$.
Solving Eq. \rf{2.23} for the Lagrangian \rf{3.29} (for $U\left(\phi\right)=1$), we get
\be{3.30}
\bar X=A\left(\frac{a}{a_{0}}\right)^{-2/\gamma}\, ,
\ee
where $A=\left(\frac{\beta^{2}}{2\gamma\left(2\gamma-1\right)}\right)^{1/\gamma}$.
Substituting the solution \rf{3.30} in Eq. \rf{3.1}, we find
\be{3.31}
V=\frac{\beta^{2}\left(1+\gamma\right)}{2\gamma\left(2\gamma-1\right)}\left(\frac{a}{a_{0}}\right)^{-2}+C\, ,
\ee
where $C$ is an integration constant. Obviously, $C$ plays the role of the cosmological constant. If the cosmological constant $\Lambda$ is already included into the model, we can put $C=0$. However, there is no need in our case to introduce in the action the cosmological constant by hand because it appears automatically in our model as a solution of the Eq. \rf{3.1}. Therefore, the $K$-essence takes the form of a two-components perfect fluid: a cosmological constant and a frustrated network of cosmic strings.
In the particular case $C=0$, the pressure \rf{3.29} and energy density \rf{1.5} of the scalar field in terms of the scale factor $a$ are given by the formulas:
\be{3.32}
\bar P=-\frac{\gamma\beta^{2}}{2\gamma\left(2\gamma-1\right)}\left(\frac{a}{a_{0}}\right)^{-2} 
\,,
\qquad \bar\varepsilon=\frac{3\gamma\beta^{2}}{2\gamma\left(2\gamma-1\right)}\left(\frac{a}{a_{0}}\right)^{-2} 
\, .
\ee
If $\gamma =1$, these formulas exactly coincide with the ones in Eq. (3.26) in Ref. \cite{Burgazli:2015mzm}. Moreover, in the case of an arbitrary $\gamma\neq 1/2$ (and $C=0$), the EoS for the $K$-essence field is $w=-\frac{1}{3}$. Therefore we can conclude
that, within the mechanical approach the non-canonical model (with
constant speed of sound) defined by Eq. \rf{3.29} for $U\left(\phi\right)=1$ exhibits equivalent dynamics to that of a canonical scalar field model \cite{Burgazli:2015mzm} even though the action is different. One further difference between these models consists on the value of the speed of sound squared: for the $K$-essence model \rf{3.29} the speed of sound squared is an arbitrary constant $c_s^2=(2\gamma-1)^{-1}$ (see Eq. \rf{1.7}) but for the canonical model $c_s^2=1$.

\

\subsection{$K$-essence model with a quartic kinetic term and an arbitrary potential}

\

Here, we study a $K$-essence model which is defined by the addition
of a quadratic term in $X$ to the kinetic energy of the canonical ($b=1$) or phantom ($b=-1$)
scalar field
\be{3.33}
P\left(X, \phi\right)=bX+cX^{2}-V\left(\phi\right)\, ,
\ee
where $b$ and $c$ are arbitrary constants. For $b=-1$ the model is called ghost condensate DE model which was originally proposed as a mean to avoid various
instabilities in phantom models \cite{Caldwell:2003vq,ArkaniHamed:2003uy}. Obviously, the case $b=0$ belongs to the model analysed on the previous subsection with $\gamma=2$ and, consequently, with constant sound speed $c_s^2=1/3$. So, we exclude this case from the analysis carried in the present subsection. Here, we also suppose that $X$ is non-negative but the parameters $b$ and $c$ may be both negative and/or positive.

Inserting \rf{3.33} in the definition of the background energy density of the $K$-essence, see Eq. \rf{1.5}, we obtain
\be{3.34}
	\bar \varepsilon\left(\bar X, \phi_c\right) =~ 
	b\bar X+3c\bar X^2 + V\left(\phi_c\right)
	\, ,
\ee
while for the speed of sound squared, we obtain from Eq. \rf{1.7}
\be{3.35}
	c_s^2
	= \frac{b+2c\bar X}{b+6c\bar X} 
	\, .
\ee
From this expression we can define a critical value $X_\textrm{crit}$,
\[
X_\textrm{crit}\equiv-b/(6c),
\]
for which the speed of sound squared $c_s^2 (\bar X)$ gets very large values.

Using the mechanical approach constraint given in Eq.~\rf{2.23}, which in this particular case reduce to an algebraic equation, we obtain
\be{3.36}
	\bar X = \frac{-b\pm |b|\sqrt{1+ 12\frac{\beta^{2}c}{b^2}\left(\frac{a_{0}}{a}\right)^{2}}}{12c}
	= \bar X_\textrm{crit}\frac{1 - \lambda\sqrt{1+ \Delta\left(\frac{a_{0}}{a}\right)^{2}}}{2}
	\, ,
\ee
where $\lambda\equiv\pm 1$ and $\Delta \equiv 12\beta^{2}c/b^2$. Therefore, we have got two branches which describes the dependence of $\bar X$ on the scale factor $a$.
In the following analysis, we prefer to eliminate $X_\textrm{crit}$ in favour of the parameters $\beta^2$ and $\Delta$. Notice that from the definition of $X_\textrm{crit}$ and $\Delta$ we have $X_\textrm{crit} = -2\beta^2/(b\Delta)$ and we can re-write the previous Eq. (\ref{3.36}) as
\be{3.37}
	\bar X =  -\frac{\beta^2}{b}\frac{1 - \lambda\sqrt{1+ \Delta\left(\frac{a_{0}}{a}\right)^{2}}}{\Delta}
	\, .
\ee
It is worth comparing this solution with the one found for the canonical scalar field model $P=X-V$ (implying $b=1$ and $c=0$) studied in a previous paper \cite{Burgazli:2015mzm}. Taking the limit $\Delta\rightarrow0$ (which is equivalent to the limit $c\rightarrow0$) of \rf{3.37}, we find that in the case of $\lambda=1$ we recover the results of \cite{Burgazli:2015mzm}, not surprising as this solution contains as a limiting case the model analysed in \rf{3.37}, while for $\lambda=-1$ the limit is not well defined, i.e., \rf{3.37} diverges as we approach $\Delta=0$. This case is again not surprising as this solution does not contain the model analysed in \cite{Burgazli:2015mzm}.

In order for the argument of the squared root in \rf{3.37} to be positive, we must have
\be{3.38}
	a^2>-\Delta\, a_0^2
	\,.
\ee
The allowed range of values for $a$ will therefore depend on the sign of the parameter $\Delta$. If $\Delta>0$, the model is well defined for arbitrarily small values of $a$, with
\be{3.39}
	\bar X(a\ll a_0)\simeq \lambda\frac{\beta^2}{b\sqrt{\Delta}}\frac{a_{0}}{a}
	\, ,\quad \Delta >0\, ,
\ee
while for $\Delta<0$ there is a minimum value of the scale factor, $a_\textrm{min}\equiv \sqrt{|\Delta|}a_0$, starting from which the model is valid. In this case, we find that as the scale factor reaches the minimum value, the kinetic variable $X$ evolves to the limiting value $\bar X(a_\textrm{min})=-\beta^2/(b\Delta)=\bar X_\textrm{crit}/2$. In order for the model to be valid at the present time, we must have $\Delta\geq-1$. It is worth noting that for the small absolute value of $|\Delta|\ll 1$, we get $a_\textrm{min}\ll a_0$. Since the parameter $c$ determines the amount of deviation from the canonical linear model,
it is natural to suppose that $c$ should be small, which can provide the smallness of $|\Delta|$.

The late time behaviour of the solution found in \rf{3.37} is given by
\be{3.40}
	\bar X(a\gg a_0) = -\frac{2\beta^2}{b}\left[
		\frac{1 - \lambda}{2\Delta}
		- \frac{\lambda}{4}\left(\frac{a_0}{a}\right)^2
		+ \frac{\lambda\Delta}{16}\left(\frac{a_0}{a}\right)^4
		+O\left(\frac{1}{a^5}\right)
	\right]
	\, .
\ee
From this expression, we find that the asymptotic behaviour of the solution depends mainly on the value of $\lambda$. If $\lambda=1$, the constant term inside the squared brackets vanishes and $\bar X\sim (a_0/a)^2$ for large values of the scale factor. On the other hand, if $\lambda=-1$ then the constant term becomes the dominant one at late time, with $\bar X(a\rightarrow+\infty)=-2\beta^2/(b\Delta)=\bar X_\textrm{crit}$.

 In addition to the parameters $\lambda$ and $\Delta$, \rf{3.37} also depends on the ratio $\beta^2/b$. Therefore, if we want to restrict our analysis to positive values of $\bar X$ we must impose:
\begin{enumerate}
	\item $\beta^2/b>0$, $\lambda=1$, and $\Delta>0$: in this case $\bar X$ is well defined for all values of the scale factor $a\in[0,+\infty)$, while taking values $\bar X\in[0,+\infty)$. In this case $\bar X_\textrm{crit}= -2\beta^2/(b\Delta)<0$.
	\item $\beta^2/b>0$, $\lambda=1$, and $\Delta<0$: in this case $\bar X$ is well defined for values of the scale factor $a\in[a_\textrm{min},+\infty)$, while taking values $\bar X\in[0,X_\textrm{crit}/2]$, where $\bar X_\textrm{crit}= -2\beta^2/(b\Delta)>0$.
	\item $\beta^2/b>0$, $\lambda=-1$, and $\Delta<0$: in this case $\bar X$ is well defined for values of the scale factor $a\in[a_\textrm{min},+\infty)$, while taking values $\bar X\in[\bar X_\textrm{crit}/2,\bar X_\textrm{crit}]$, where $\bar X_\textrm{crit}= -2\beta^2/(b\Delta)>0$.
	\item $\beta^2/b<0$, $\lambda=-1$, and $\Delta>0$: in this case $\bar X$ is well defined for values of the scale factor $a\in[0,+\infty)$, while taking values $\bar X\in[X_\textrm{crit},+\infty)$, where $\bar X_\textrm{crit}= -2\beta^2/(b\Delta)>0$.
\end{enumerate}
As we will see below, only the first two cases are free of divergences in the physical quantities $\bar \varepsilon,\, \bar P$ of the $K$-essence as $a\rightarrow+\infty$ (or, equivalently, as $\bar X \to \bar X_{\mathrm{crit}}$).

Taking into account the mechanical approach constraint \rf{3.36}, we can re-write the pressure and energy density of the $K$-essence field, Eqs. \rf{3.33} and \rf{3.34}, as
\ba{3.41}
	&{}&\bar P = -\frac{2\beta^2}{\Delta} \frac{\bar X}{\bar X_\textrm{crit}}\left[1 - \frac{1}{6}\frac{\bar X}{\bar X_\textrm{crit}}\right] - V\left(\phi_c\right)
		=\frac{\beta^2}{12}
	\left[
		\left(\frac{a_{0}}{a}\right)^{2}
		+ 10\frac{\lambda\sqrt{1+ \Delta\left(\frac{a_{0}}{a}\right)^{2}} -1}{\Delta}
	\right]
	 - V\left(\phi_c\right)\, ,\nonumber \\
	\\
	\label{3.42}
	&{}&\bar \varepsilon = -\frac{2\beta^2}{\Delta} \frac{\bar X}{\bar X_\textrm{crit}}\left[1 - \frac{1}{2}\frac{\bar X}{\bar X_\textrm{crit}}\right] + V\left(\phi_c\right)
	=
	\frac{\beta^2}{4}
	\left[
		\left(\frac{a_{0}}{a}\right)^{2}
		+ 2\frac{\lambda\sqrt{1+ \Delta\left(\frac{a_{0}}{a}\right)^{2}} -1}{\Delta}
	\right]
	+ V\left(\phi_c\right)
	\, .\nonumber \\
\ea
Hence,
\be{3.43}
	\bar \varepsilon + \bar P =~
	\frac{\beta^2}{3}
	\left[
		\left(\frac{a_{0}}{a}\right)^{2}
		+ 4\frac{\lambda\sqrt{1+ \Delta\left(\frac{a_{0}}{a}\right)^{2}} - 1}{\Delta}
	\right]
	\, .
\ee
To evaluate the expressions \rf{3.41}-\rf{3.43} at the present time, we should simply put $a=a_0$ in these formulas.

According to experimental data (see e.g. \cite{Planck2013}), for a given model to account for the current cosmic acceleration, the condition { that $|\bar \varepsilon + \bar P|$ is very small} must hold (at least from an effective point of view as it is the case for example on modified theories of gravity (see \cite{JoaoM} and references therein)). Can the considered model satisfy this condition starting at $a<a_0$? To answer this question, we consider the case of small values of $\Delta$, i.e.  $|\Delta| \ll 1$. As we mentioned above, this is the physically reasonable case. We also consider scale factors such that the condition $|\Delta|a^2_0/a^2\ll 1$ is fulfilled; i.e. scale factors larger than $a_{\textrm{min}}$. Obviously, the condition $|\Delta|a^2_0/a^2\ll 1$ is compatible with  $a<a_0$ as we are imposing $|\Delta| \ll 1$. Then, Eq. \rf{3.43}
can be expanded as (for $a\to +\infty$)
\ba{3.44}
1.\; \lambda = +1:\quad \bar \varepsilon + \bar P &=& \beta^2\left(\frac{a_0}{a}\right)^2 + O\left(\Delta (a_0/a)^4\right)\; \to \; 0   \, ,\\
\label{3.45}
2.\; \lambda = -1:\quad\bar \varepsilon + \bar P &=& -\frac{\beta^2}{3}\left[\frac{8}{\Delta}+\left(\frac{a_0}{a}\right)^2\right]
+ O\left(\Delta (a_0/a)^4\right)\; \to \; -\frac{\beta^2}{3}\frac{8}{\Delta} \, .
\ea
These expressions demonstrate that for sufficiently small values of $\beta^2$, we can achieve the desired condition, starting from $a<a_0$, in the first case
$\lambda=+1$. However, this is quite problematic for the second case $\lambda=-1$.

To obtain the dependence of the $K$-essence scalar field potential $V$ on the scale factor,  we insert Eq. \rf{3.37} in Eq. \rf{3.1} and find the following expression for the potential $V(a)$:
\be{3.46}
	V\left(a\right)= V_\infty + \frac{\beta^2}{4}\left[
		\left(\frac{a_0}{a}\right)^2
		+ 14\lambda\frac{\sqrt{1+\Delta\left(\frac{a_0}{a}\right)^2} - 1}{\Delta}
		-16\frac{1}{\Delta}\log\left(\frac{1+\sqrt{1+\Delta}}{2}\frac{1+ \lambda\sqrt{1+\Delta\left(\frac{a_0}{a}\right)^2}}{1 + \lambda\sqrt{1+\Delta}}\right)
	\right]
	\,.
\ee
where $V_\infty$ is an integration constant. This constant is connected with the present time value $V_0$ as follows:
\be{3.47}
	V_0 = V_{\infty}
	+ \frac{\beta^2}{4}\left[
		1
		+ 14\lambda\frac{\sqrt{1+\Delta}-1}{\Delta}
		- 16\frac{1}{\Delta}\log\left(\frac{1+\sqrt{1+\Delta}}{2}\right)
	\right]
	\, .
\ee
In the case $|\Delta|\ll 1$ and $\lambda=+1$, this expression is reduced to $V_{\infty}=V_0-\beta^2 +O(\Delta)$.
We shall see below that $V_{\infty}$ corresponds to a cosmological constant.  It can be put to zero by fine tuning
$V_0$ and the parameters of the model. It can be also easily seen that in the case $\lambda=-1$ the potential \rf{3.46} diverges if $a\to +\infty$ (equivalently, if $\bar X \to \bar X_{\mathrm{crit}}$).

Substituting \rf{3.46} in Eqs. \rf{3.41} and \rf{3.42}, we finally obtain
\ba{3.48}
	\bar P &=&
		- V_\infty
	- \frac{\beta^2}{6}\left[
		\left(\frac{a_0}{a}\right)^2
		+ 5\frac{1-\lambda}{\Delta}
		+ 16\lambda\frac{\sqrt{1+\Delta\left(\frac{a_0}{a}\right)^2} - 1}{\Delta}\right.\nn\\
		&-&\left. 24\frac{1}{\Delta}\log\left(\frac{1+\sqrt{1+\Delta}}{2}\frac{1+ \lambda\sqrt{1+\Delta\left(\frac{a_0}{a}\right)^2}}{1 + \lambda\sqrt{1+\Delta}}\right)
	\right]
\ea
and
\ba{3.49}
	\bar\varepsilon &=&
		V_\infty
	+ \frac{\beta^2}{2}\left[
		\left(\frac{a_0}{a}\right)^2
		- \frac{1-\lambda}{\Delta}
		+ 8\lambda\frac{\sqrt{1+\Delta\left(\frac{a_0}{a}\right)^2} - 1}{\Delta}\nn\right.\\
		&-&\left.8\frac{1}{\Delta}\log\left(\frac{1+\sqrt{1+\Delta}}{2}\frac{1+ \lambda\sqrt{1+\Delta\left(\frac{a_0}{a}\right)^2}}{1 + \lambda\sqrt{1+\Delta}}\right)
	\right]
	\, .
\ea
In the case $\lambda =-1$, these expressions have also logarithmic divergences if $a\to +\infty$ ($\bar X \to \bar X_{\mathrm{crit}}$).

It also useful to obtain the speed of sound squared as a function of the scale factor. From Eqs. \rf{3.35} and \rf{3.37} we get
\be{3.50}
	c_s^2 = \frac{1}{3} + \frac{2}{3}\frac{1}{1-X/X_\textrm{crit}}
	=\frac{1}{3} + \frac{4}{3}\frac{1}{1 + \lambda \sqrt{1 + \Delta\left(\frac{a_0}{a}\right)^2}}
	\, .
\ee

Let us now take a closer look at the late-time behaviour of $c_s^2$, $V$, $\bar\varepsilon$, $\bar P$, $\bar \varepsilon+P$ for this model of $K$-essence in the limit of very large values of the scale factor, i.e., $a\gg a_0$. As we have seen above, the limiting behaviour of the model depends on whether $\lambda=\pm1$. We start by analysing the behaviour of these quantities for $\lambda=+1$. Expanding \rf{3.43}, \rf{3.46}, \rf{3.48}, \rf{3.49}, and \rf{3.50} around $1/a=0$, we find
\ba{3.51}
	{c_s^2}\left(a\gg a_0\right)  &\sim& 1 - \frac{\Delta}{6}\left(\frac{a_0}{a}\right)^2 + \frac{\Delta^2}{12}\left(\frac{a_0}{a}\right)^4 + O\left(\frac{1}{a^5}\right)
	\, ,\\
	\label{3.52}V\left(a\gg a_0\right) &\sim& V_\infty + \beta^2\left(\frac{a_0}{a}\right)^2 - \frac{\beta^2\Delta}{16}\left(\frac{a_0}{a}\right)^4 + O\left(\frac{1}{a^5}\right)
	\, ,\\
	\label{3.53}\bar \varepsilon(a\gg a_0)  &\sim& V_\infty + \frac{3\beta^2}{2}\left(\frac{a_0}{a}\right)^2 - \frac{\beta^2\Delta}{8}\left(\frac{a_0}{a}\right)^4 + O\left(\frac{1}{a^5}\right)
	\, ,\\
	\label{3.54}\bar P(a\gg a_0)  &\sim& - V_\infty - \frac{\beta^2}{2}\left(\frac{a_0}{a}\right)^2 - \frac{\beta^2\Delta}{24}\left(\frac{a_0}{a}\right)^4 + O\left(\frac{1}{a^5}\right)
	\, ,\\
	\label{3.55}(\bar\varepsilon+\bar P)(a\gg a_0)  &\sim& \beta^2\left(\frac{a_0}{a}\right)^2 - \frac{\beta^2\Delta}{6}\left(\frac{a_0}{a}\right)^4 + O\left(\frac{1}{a^5}\right)
	\, .
\ea
Therefore, for $\lambda=+1$ the $K$-essence scalar field behaves as a three component perfect fluid: $a$ cosmological constant, $a$ frustrated network of cosmic strings and dark radiation.
Notice that in the limiting case $\Delta\equiv 0$ we restore the results of our previous paper
\cite{Burgazli:2015mzm}. In fact, if we set $\Delta=0$ in the previous expressions we recover the exact results of that paper, where it was imposed $V_\infty=V_0-\beta^2=0$. It is worth noting also that the asymptotic formulas \rf{3.51}-\rf{3.55} also take place in the case $|\Delta|\ll 1$ for the scale factors satisfying the condition $|\Delta|a^2_0/a^2\ll 1$, that is for $a\ll a_0$.

Similarly, in the case $\lambda=-1$, we obtain instead
\ba{3.56}
	{c_s^2}\left(a\gg a_0\right)   &\sim&
	-\frac{8}{3\Delta}\left(\frac{a_0}{a}\right)^{-2}
	-\frac{1}{3}
	 + \frac{\Delta}{6} \left(\frac{a_0}{a}\right)^{2}
	 - \frac{\Delta^2}{12}\left(\frac{a_0}{a}\right)^{4}
	 + O\left(\frac{1}{a^5}\right)
	\, ,\\
	\label{3.57}V\left(a\gg a_0\right)  &\sim&
	V_\infty
	+\frac{4\beta^2}{\Delta}\log\left[\frac{4}{(1+\sqrt{1+\Delta})^2}\left(\frac{a}{a_0}\right)^{2} \right]\nn\\
	&-& \frac{\beta^2}{2}\left(\frac{a_0}{a}\right)^2
	+ \frac{\beta^2\Delta}{16}\left(\frac{a_0}{a}\right)^4
	+ O\left(\frac{1}{a^5}\right)
	\, ,\\
	\label{3.58}\bar\varepsilon\left(a\gg a_0\right)  &\sim&
	V_\infty
	-\frac{\beta^2}{\Delta}
	+\frac{4\beta^2}{\Delta}\log\left[\frac{4}{(1+\sqrt{1+\Delta})^2}\left(\frac{a}{a_0}\right)^{2} \right]\nn\\
	&-& \frac{\beta^2}{2}\left(\frac{a_0}{a}\right)^2
	+ \frac{\beta^2\Delta}{8}\left(\frac{a_0}{a}\right)^4
	+ O\left(\frac{1}{a^5}\right)
	\, ,\\
	\label{3.59}\bar P\left(a\gg a_0\right)  &\sim&
	-V_\infty
	-\frac{5\beta^2}{3\Delta}
	-\frac{4\beta^2}{\Delta}\log\left[\frac{4}{(1+\sqrt{1+\Delta})^2}\left(\frac{a}{a_0}\right)^{2} \right]\nn\\
	&+& \frac{\beta^2}{6}\left(\frac{a_0}{a}\right)^2
	+ \frac{\beta^2\Delta}{24}\left(\frac{a_0}{a}\right)^4
	+ O\left(\frac{1}{a^5}\right)
	\, ,\\
	\label{3.60}(\bar\varepsilon+\bar P)(a\gg a_0)  &\sim&
	-\frac{8\beta^2}{3\Delta}
	- \frac{\beta^2}{3}\left(\frac{a_0}{a}\right)^2
	+ \frac{\beta^2\Delta}{6}\left(\frac{a_0}{a}\right)^4
	+ O\left(\frac{1}{a^5}\right)
	\, .
\ea
We, therefore, find that, for $\lambda=-1$, this model of K-essence evolves asymptotically to a Little Sibling of the Big Rip event \cite{Bouhmadi-Lopez:2014cca,Imanol}; i.e. the Hubble rate and the scale factor blow up but the cosmic time derivative of the Hubble rate does not. In the distant future, the energy density and the pressure, driven by the potential, diverge with $\sim\log(a)$ while their sum remains constant. As it follows from Eq. \rf{3.58}, the energy density $\bar\varepsilon$ is always positive for large values of $a$ if and only if $\beta^2/\Delta=b^2/(12 c)>0$; i.e. $c>0$. Therefore, we should demand $c>0$, in the case $\lambda=-1$, to avoid having negative energies in the future. The squared speed of sound also diverges with $\sim a^2$. A simple analysis shows that
it is possible to achieve the late-time cosmic acceleration in this case.

It makes sense to introduce also a dimensionless energy density for the $K$-essence scalar field, i.e. we simply redefine the terms involved on the energy density of the $K$-essence field as follows (see Eq. \rf{3.49}):
\ba{3.61}
	&{}&\Omega_\textrm{essence} \equiv~ \frac{\kappa^2\bar \varepsilon}{3H_{0}^{2}}
	\\&=&
	\Omega_\textrm{essence,0}
	+ \frac{\beta^2_*}{2}\left[
		\left(\frac{a_0}{a}\right)^2-1
		+ 8\lambda\frac{\sqrt{1+\Delta\left(\frac{a_0}{a}\right)^2} - \sqrt{1+\Delta}}{\Delta}
		-8\frac{1}{\Delta}\log\left(\frac{1+ \lambda\sqrt{1+\Delta\left(\frac{a_0}{a}\right)^2}}{1 + \lambda\sqrt{1+\Delta}}\right)
	\right]\nn
		\, ,
\ea
where $\Omega_\textrm{essence,0}=\kappa^2\bar\varepsilon_0/(3H_{0}^{2})$ is its value at the present time and we have absorbed $\beta$ into the dimensionless parameter ${\beta}_*=\sqrt{\kappa^2/(3H_{0}^{2})}\beta$. We can also
express the EoS as a function of the scale factor and in terms of
the dimensionless parameters $\left(\lambda,\,\Delta,\,\beta_*,\,\Omega_\textrm{essence,0}\right)$:
\ba{3.62}
&{}&w = \frac{\bar P}{\bar\varepsilon}  = -1 + \frac{\bar \varepsilon+\bar P}{\bar\varepsilon}
\\&=&
	-1
	+
	\frac{
		\frac{\beta_*^2}{3}
		\left[
			\left(\frac{a_{0}}{a}\right)^{2}
			+ 4\frac{\lambda\sqrt{1+ \Delta\left(\frac{a_{0}}{a}\right)^{2}}-1}{\Delta}
		\right]
	}
	{
		\Omega_\textrm{essence,0}
		+ \frac{\beta_*^2}{2}\left[
			\left(\frac{a_0}{a}\right)^2 - 1
			+ 8\lambda\frac{\sqrt{1+\Delta\left(\frac{a_0}{a}\right)^2} - \sqrt{1+\Delta}}{\Delta}
			-8\frac{1}{\Delta}\log\left(\frac{1+ \lambda\sqrt{1+\Delta\left(\frac{a_0}{a}\right)^2}}{1 + \lambda\sqrt{1+\Delta}}\right)
		\right]
	}\nn
	\, .
\ea
At the present time, this equation reads
\be{3.63}
	w_0 = -1 +
	\frac{\beta_*^2}{3\Omega_\textrm{essence,0}}
	\left[
			1
			+ 4\frac{\lambda\sqrt{1+ \Delta}-1}{\Delta}
		\right]
	\, .
\ee
In the most interesting and natural case $\lambda=+1$ and $|\Delta|\ll 1$, we get
\be{3.64}
w_0 \approx -1 +
	\frac{\beta_*^2}{\Omega_\textrm{essence,0}}\, .
\ee
According to the current observations \cite{Planck2013}, the parameter $w$ of the dark energy EoS must be rather close to $-1$.
Therefore, we should constrain the parameters of the model as follows:
$|\beta_*^2/\Omega_\textrm{essence,0}|=|\beta^2/\bar\varepsilon_0|\ll 1$. Taking into account that in the considered case
$\bar\varepsilon_0 \approx V_0+\beta^2/2$, we get the condition $|V_0/\beta^2|>>1$. From \rf{3.47} we have $V_0 \approx V_{\infty} +\beta^2$. Hence,  to describe the current cosmic acceleration in the case $\lambda=+1$ and $|\Delta|\ll 1$, we should demand $|V_{\infty}/\beta^2|>>1$. 


\section{\label{sec:5}Conclusion}

In the present paper, we have studied the current Universe which we have assumed to be filled with dust, radiation and a $K$-essence scalar field playing the role of dark energy.
We have considered the Universe at the late stage of its evolution and deep inside of the cell of uniformity.
At such scales the Universe is highly inhomogeneous: we can clearly see here the discrete distributed inhomogeneities
in the form of galaxies and groups of the galaxies. These inhomogeneities represent the dust component of matter in the Universe.
We include the $K$-essence component to provide an explanation of the late-time acceleration of the Universe.
We have considered the $K$-essence in a very specific coupled form \cite{coupled}. 
The point is that at the late stage of the Universe evolution and deep inside the cell of uniformity, the inhomogeneities (e.g. galaxies) have non-relativistic peculiar velocities and the
mechanical approach (which is an ideal technique to study the scalar perturbations and to obtain the gravitational potential of the inhomogeneities, being the peculiar velocities neglected) is an adequate tool to describe cosmological models \cite{Eingorn:2012jm,EZcosm2,EKZ2}. We have
studied the possibility that the fluctuations of the energy density and pressure of the $K$-essence scalar field are concentrated around the galaxies and the group of the galaxies. Therefore, they also have non-relativistic peculiar velocities. Then, if such coupled $K$-essence scalar fields exist,
can they provide the late cosmic acceleration? Hence, the main objective of the article was to address these two problems; i.e. what are the $K$-essence models (at least among those analysed on this paper) that are compatible with the cosmological scalar perturbations within the mechanical approach and at the same time are able to describe the current speed up of the Universe?

First, we have shown that for the $K$-essence to be coupled, it should satisfy the master Eq. \rf{2.23}. Under such a condition,
the fluctuations of the energy density and pressure of the $K$-essence are concentrated around the inhomogeneities screening their gravitational potentials (see Eqs. \rf{2.26} and \rf{2.27}).  Then, we have considered a number of particular types of $K$-essence fields to study their possibility to accelerate the late-time Universe. These models are: (i) the pure kinetic $K$-essence, (ii) the constant speed of sound $K$-essence and (iii) the $K$-essence model with the kinetic term $bX+cX^2$. We have shown that if they are coupled, all these $K$-essence
scalar fields take the form of multicomponent perfect fluids where one of the component is the cosmological constant. Therefore, all these models can result in the late-time cosmic acceleration. In this case, observations provide restrictions on the parameters of the models.

 To conclude, we would like to mention about an interesting generalization of our model that consists in investigating the coupling between dark matter and dark energy which can be modelled with scalar fields (see e.g. \cite{BCP}). Within the scope of the mechanical approach, we have already considered the interaction between dark matter and dark energy for a more phenomenological model \cite{BLMZ}. It is of course of interest to investigate this problem in the mechanical approach in the case of interacting scalar fields. { This will modify the scheme of our approach as the conservation of the energy momentum tensor of dark matter and dark energy will be affected and a careful and detailed analysis will be required.} We will tackle this issue in the near future.

\section*{Acknowledgements}

The work of MBL is supported by the Portuguese Agency "Funda\c{c}\~ao para a Ci\^encia e Tecnologia" through an Investigador FCT Research contract, with reference IF/01442/2013/ CP1196/CT0001. She also wishes to acknowledge the partial support from the Basque government Grant No. IT592-13 (Spain) and FONDOS FEDER under grant FIS2014-57956-P (Spanish government). This research work is supported by the Portuguese grant UID/MAT/00212/2013. J. Morais is thankful to UPV/EHU for a PhD fellowship and UBI for hospitality during the completion of part of this work and acknowledges the support from the Basque government Grant No. IT592-13 (Spain) and Fondos FEDER, under grant FIS2014-57956-P (Spanish Government). A.Zh. acknowledges the hospitality of UBI during his visit in 2015 during the completion of part of this work.  SK acknowledges for the support of grant SFRH/BD/51980/2012 from the Portuguese Agency ``Funda\c{c}\~ao para a Ci\^encia e Tecnologia''.



\begin{thebibliography}{99}

\bibitem{SN1}
A.G. Riess et al., {\em Observational evidence from supernovae for an accelerating Universe and a cosmological constant}, Astron. J. {\bf 116} (1998) 1009
 [\href{http://arxiv.org/abs/astro-ph/9805201}{astro-ph/9805201v1}].
 
\bibitem{SN2}
S. Perlmutter et al., {\em Measurements of Omega and Lambda from 42 high-redshift supernovae}, Astrophys. J. {\bf 517} (1999) 565 
 [\href{http://arxiv.org/abs/astro-ph/9812133}{astro-ph/9812133v1}].
 
\bibitem{quintess1}
P.G. Ferreira, M. Joyce, {\em Structure formation with a self-tuning scalar field}, Phys. Rev. Lett. {\bf 79} (1997) 4740  
 [\href{http://arxiv.org/abs/astro-ph/9707286}{astro-ph/9707286v1}].
 
\bibitem{quintess1a}
R.R. Caldwell, R. Dave and P.J. Steinhardt, {\em Cosmological Imprint of an Energy Component with General Equation of State}. Phys. Rev. Lett. {\bf 80} (1998) 1582  
 [\href{http://arxiv.org/abs/astro-ph/9708069}{astro-ph/9708069v2}].
 
\bibitem{quintess2}
L. Wang, P.J. Steinhardt, {\em Cluster abundance constraints for cosmological models with a timevarying, spatially inhomogeneous energy component with negative pressure}, Astrophys. J. {\bf 508} (1998) 483  
 [\href{http://arxiv.org/abs/astro-ph/9804015}{astro-ph/9804015v1}].
 
\bibitem{quintess3}
I.~Zlatev, L. Wang, P.J. Steinhardt, {\em Quintessence, cosmic coincidence, and the cosmological constant}, Phys. Rev. Lett. {\bf 82} (1999) 896  
 [\href{http://arxiv.org/abs/astro-ph/9807002}{astro-ph/9807002v2}].
 
\bibitem{phantom1}
R.R. Caldwell, {\em A phantom menace? Cosmological consequences of a dark energy component with super-negative equation of state}, Phys. Lett. B {\bf 545} (2002) 23  
 [\href{http://arxiv.org/abs/astro-ph/9908168}{astro-ph/9908168v2}].
 
\bibitem{phantom2}
S.M. Carroll, M. Hoffman, M. Trodden, {\em Can the dark energy equation-of-state parameter w be less than -1?} Phys. Rev. D. {\bf 68} (2003) 023509  
 [\href{http://arxiv.org/abs/astro-ph/0301273}{astro-ph/0301273v2}].
 
\bibitem{quintom}
Y.-F. Cai, E. N. Saridakis, M. R. Setare and J.-Q. Xia,
{\em Quintom Cosmology: Theoretical implications and observations}, Phys. Rept. {\bf 493} (2010) 1 
 [\href{http://arxiv.org/abs/0909.2776}{arXiv:0909.2776v2}].

\bibitem{Dolgov}
A.D. Dolgov, {\em Cosmic antigravity},
 [\href{http://arxiv.org./abs/1206.3725}{arXiv:1206.3725v1}].
 
\bibitem{zhuk1996}
A. Zhuk,
{\em Integrable scalar field multi-dimensional cosmologies}, Class. Quant. Grav. {\bf 13} (1996) 2163.

\bibitem{ZBG2}
M.~Bouhmadi-L\'opez, P.~F.~Gonz\'alez-D\'iaz and A.~Zhuk,
{\em On New gravitational instantons describing creation of brane worlds}, Class.\ Quant.\ Grav.\  {\bf 19} (2002) 4863  
 [\href{http://arxiv.org/abs/hep-th/0208226}{hep-th/0208226v1}].
 
\bibitem{ZBG}
M. Bouhmadi-L\'opez, P. F. Gonz\'alez-D\'iaz, A. Zhuk,
{\em Topological defect brane-world models}, Gravitation and Cosmology {\bf 8} (2002) 285  
 [\href{http://arxiv.org/abs/hep-th/0207170}{hep-th/0207170v2}].
 
\bibitem{ArmendarizPicon:1999rj}
C.~Armendariz-Picon, T.~Damour, and V.~F. Mukhanov, {\em k - inflation}, Phys. Lett. B {\bf 458} (1999) 209 
[\href{http://arxiv.org/abs/hep-th/9904075}{hep-th/9904075v1}].

\bibitem{Garriga:1999vw}
J.~Garriga and V.~F. Mukhanov, {\em Perturbations in k-inflation},Phys. Lett. B {\bf 458} (1999) 219 
[\href{http://arxiv.org/abs/hep-th/9904176}{hep-th/9904176v1}].

\bibitem{book}
B. Novosyadlyj, V. Pelykh, Yu. Shtanov and A. Zhuk,
 Dark energy and dark matter in the Universe: in three volumes.
Editor V. Shulga. - Vol. 1.  Dark Energy: observational evidence and
theoretical models. - Kiev: Akademperiodyka, 2013,  380 pages;
 [\href{http://arxiv.org./abs/1502.04177}{arXiv:1502.04177v1}].
 
\bibitem{Ein1}
M. Eingorn,
{\em First-order Cosmological Perturbations Engendered by Point-like Masses},
Astrophys.\ J.\  {\bf 825} (2016) 84
 [\href{http://arxiv.org./abs/1509.03835}{arXiv:1509.03835v3}].

\bibitem{Eingorn:2012jm}
M.~Eingorn and A.~Zhuk, {\em Hubble flows and gravitational potentials in
  observable Universe},  JCAP {\bf 1209} (2012) 026  
[\href{http://arxiv.org/abs/1205.2384}{arXiv:1205.2384v2}].

\bibitem{EZcosm2}
M. Eingorn and A. Zhuk, {\em Remarks on mechanical approach to observable Universe}, JCAP {\bf 05} (2014) 024  
[\href{http://arxiv.org/abs/1309.4924}{arXiv:1309.4924v2}].

\bibitem{EKZ2}
M. Eingorn, A. Kudinova, A. Zhuk,
{\em Dynamics of astrophysical objects against the cosmological background}, JCAP {\bf 04} (2013) 010 
[\href{http://arxiv.org/abs/1211.4045}{arXiv:1211.4045v2}].

\bibitem{coupled}
A. Zhuk,
{\em Perfect fluids coupled to inhomogeneities in the late Universe}, Gravitation and Cosmology {\bf 22} (2016) 159
[\href{http://arxiv.org/abs/1601.01939}{arXiv:1601.01939v1}].

\bibitem{Burgazli:2013qy}
A.~Burgazli, M.~Eingorn, and A.~Zhuk, {\em Rigorous theoretical constraint on
  constant negative EoS parameter $\omega $ and its effect for the late
  Universe}, Eur. Phys. J. C {\bf 75} (2015) 118 
[\href{http://arxiv.org/abs/1301.0418}{arXiv:1301.0418v3}].

\bibitem{ENZ1}
M. Eingorn, J. Nov$\mathrm{\acute{a}}$k and A. Zhuk, {\em $f(R)$ gravity: scalar perturbations in the late Universe}, 
Eur. Phys. J. C {\bf 74} (2014) 3005 
[\href{http://arxiv.org/abs/1401.5410}{arXiv:1401.5410v2}].

\bibitem{Laslo2}
M. Brilenkov, M. Eingorn, L. Jenkovszky and A.~Zhuk, {\em Scalar perturbations in cosmological models with quark nuggets}, Eur. Phys. J. C {\bf 74} (2014) 3011 
[\href{http://arxiv.org/abs/1310.4540}{arXiv:1310.4540v2}].

\bibitem{Bouhmadi-Lopez:2015oxa}
M.~Bouhmadi-L\'opez, M.~Brilenkov, R.~Brilenkov, J.~Morais, and A.~Zhuk,
{\em Scalar perturbations in the late Universe: viability of the Chaplygin gas
  models}, JCAP {\bf 1512}  no.12 (2015) 037 
[\href{http://arxiv.org/abs/1509.06963}{arXiv:1509.06963v2}].


\bibitem{Burgazli:2015mzm}
A.~Burgazli, A.~Zhuk, J.~Morais, and K.~S. Kumar, {\em Scalar fields in the late
  Universe: The mechanical approach},
[\href{http://arxiv.org/abs/1512.03819v1}{arXiv:1512.03819v1}].

\bibitem{CPL}
\"{O}. Akarsu, M. Bouhmadi--L\'opez, M. Brilenkov, R. Brilenkov, M. Eingorn, A. Zhuk,
 {\em Are CPL models compatible with the history of our Universe?}, JCAP {\bf 07} (2015) 038  
[\href{http://arxiv.org/abs/1502.04693}{arXiv:1502.04693v2}].

\bibitem{Malik:2008im}
K.~A. Malik and D.~Wands, {\em Cosmological perturbations}, Phys. Rept. {\bf 475} (2009) 1 
[\href{http://arxiv.org/abs/0809.4944}{arXiv:0809.4944v2}].

\bibitem{Chiba:1999ka}
T.~Chiba, T.~Okabe, and M.~Yamaguchi, {\em Kinetically driven quintessence},
   Phys. Rev. D {\bf 62} (2000) 023511 
[\href{http://arxiv.org/abs/astro-ph/9912463}{astro-ph/9912463v2}].

\bibitem{ArmendarizPicon:2000dh}
C.~Armendariz-Picon, V.~F. Mukhanov, and P.~J. Steinhardt, {\em A Dynamical
  solution to the problem of a small cosmological constant and late time cosmic
  acceleration}, Phys. Rev. Lett. {\bf 85} (2000) 4438 
[\href{http://arxiv.org/abs/astro-ph/0004134}{astro-ph/0004134v1}].

\bibitem{Chimento:2003zf}
L.~P. Chimento and A.~Feinstein, {\em Power-law expansion in k-essence
  cosmology},  Mod. Phys. Lett. A {\bf 19} (2004) 761 
[\href{http://arxiv.org/abs/astro-ph/0305007}{astro-ph/0305007v2}].

\bibitem{Chimento:2003ta}
L.~P. Chimento, {\em Extended tachyon field, Chaplygin gas and solvable k-essence
  cosmologies}, Phys. Rev. D {\bf 69} (2004) 123517 
[\href{http://arxiv.org/abs/astro-ph/0311613}{astro-ph/0311613v2}].

\bibitem{dePutter:2007ny}
R.~de~Putter and E.~V. Linder, {\em Kinetic k-essence and Quintessence},
   Astropart. Phys. {\bf 28} (2007) 263 
[\href{http://arxiv.org/abs/0705.0400}{arXiv:0705.0400v2}].

\bibitem{Scherrer:2004au}
R.~J. Scherrer, {\em Purely kinetic k-essence as unified dark matter}, Phys. Rev. Lett.
  {\bf 93} (2004) 011301 
[\href{http://arxiv.org/abs/astro-ph/0402316}{astro-ph/0402316v3}].

\bibitem{BouhmadiLopez:2010vi}
  M.~Bouhmadi-L\'opez and L.~P.~Chimento,
  {\em k-essence in the DGP brane-world cosmology},
  Phys.\ Rev.\ D {\bf 82} (2010) 103506 
[\href{http://arxiv.org/abs/1007.4141}{arXiv:1007.4141v2}].

\bibitem{BouhmadiLopez:2006fu} 
  M.~Bouhmadi-L\'opez, P.~F.~Gonz\'alez-D\'iaz and P.~Mart\'in-Moruno,
  {\em Worse than a big rip?},
  Phys.\ Lett.\ B {\bf 659} (2008) 1  
[\href{http://arxiv.org/abs/gr-qc/0612135}{gr-qc/0612135v2}].

\bibitem{BouhmadiLopez:2007qb} 
  M.~Bouhmadi-L\'opez, P.~F.~Gonz\'alez-D\'iaz and P.~Mart\'in-Moruno,
  {\em On the generalised Chaplygin gas: Worse than a big rip or quieter than a sudden singularity?},
  Int.\ J.\ Mod.\ Phys.\ D {\bf 17} (2008) 2269 
[\href{http://arxiv.org/abs/0707.2390}{arXiv:0707.2390v2}].

\bibitem{FNCS}
M.~Bouhmadi-L\'opez, P.~Chen, Y.~C.~Huang and Y.~H.~Lin,
{\em  Slow-roll inflation preceded by a topological defect phase \`a la Chaplygin gas},
  Phys.\ Rev.\ D {\bf 87} (2013) no.10,  103513
[\href{http://arxiv.org/abs/1212.2641}{arXiv:1212.2641v2}].

\bibitem{Amendola:2004qb}
L.~Amendola, {\em Phantom energy mediates a long-range repulsive force},
  Phys. Rev. Lett. {\bf 93} (2004) 181102 
[\href{http://arxiv.org/abs/hep-th/0409224}{hep-th/0409224v2}]

\bibitem{Piazza:2004df}
F.~Piazza and S.~Tsujikawa, {\em Dilatonic ghost condensate as dark energy},
 JCAP {\bf 0407} (2004) 004 
[\href{http://arxiv.org/abs/hep-th/0405054}{hep-th/0405054v2}].

\bibitem{delaMacorra:2002du}
A.~de~la Macorra and H.~H. Vucetich, {\em Causality, stability and sound speed in
  scalar field models},
[\href{http://arxiv.org/abs/astro-ph/0212302}{astro-ph/0212302v2}].

\bibitem{Erickson:2001bq}
J.~K. Erickson, R.~R. Caldwell, P.~J. Steinhardt, C.~Armendariz-Picon, and
  V.~F. Mukhanov, {\em Measuring the speed of sound of quintessence},
Phys. Rev. Lett. {\bf 88} (2002) 121301 
[\href{http://arxiv.org/abs/astro-ph/0112438}{astro-ph/0112438v1}].

\bibitem{BS}
M. Bucher and D.N. Spergel, {\em Is the Dark Matter a Solid?} Phys. Rev. D {\bf 60} (1999) 043505 
 [\href{http://arxiv.org/abs/astro-ph/9812022}{astro-ph/9812022v3}]. 

\bibitem{BCCM}
R.A. Battye, B. Carter, E. Chachoua and A. Moss, {\em Rigidity and stability of cold dark solid universe model}, Phys. Rev. D {\bf 72} (2005) 023503 
[\href{http://arxiv.org/abs/hep-th/0501244}{hep-th/0501244v2}].

\bibitem{Conversi}
L. Conversi, A. Melchiorri, L. Mersini and J. Silk, {\em Are Domain Walls ruled out?} Astropart. Phys. {\bf 21} (2004) 443  
 [\href{http://arxiv.org/abs/astro-ph/0402529}{astro-ph/0402529v1}].

\bibitem{Kamenshchik:2001cp}
 A.~Y.~Kamenshchik, U.~Moschella and V.~Pasquier,
 {\em An Alternative to quintessence},
 Phys.\ Lett.\ B {\bf 511} (2001) 265 
 [\href{http://arxiv.org/abs/gr-qc/0103004}{gr-qc/0103004v2}].

\bibitem{Bilic:2001cg}
 N.~Bili{\'c}, G.~B.~Tupper and R.~D.~Viollier,
 {\em Unification of dark matter and dark energy: The Inhomogeneous Chaplygin gas},
 Phys.\ Lett.\ B {\bf 535} (2002) 17 
 [\href{http://arxiv.org/abs/astro-ph/0111325}{astro-ph/0111325v2}].

\bibitem{Bento:2002ps}
M.~C. Bento, O.~Bertolami, and A.~A. Sen, {\em Generalized Chaplygin gas,
  accelerated expansion and dark energy matter unification}, Phys. Rev. D {\bf 66} (2002) 043507 
[\href{http://arxiv.org/abs/gr-qc/0202064}{gr-qc/0202064v2}].

\bibitem{BouhmadiLopez:2004me}
 M.~Bouhmadi-L\'opez and J.~A.~Jim\'enez Madrid,
 {\em Escaping the big rip?}, JCAP {\bf 0505} (2005) 005 
 [\href{http://arxiv.org/abs/astro-ph/0404540}{astro-ph/0404540v1}].

\bibitem{Sergijenko:2014pwa}
O.~Sergijenko and B.~Novosyadlyj, {\em Sound speed of scalar field dark energy:
  weak effects and large uncertainties}, Phys. Rev. D {\bf 91} no.~8 (2015) 083007 
[\href{http://arxiv.org/abs/1407.2230v2}{arXiv:1407.2230v2}].

\bibitem{Caldwell:2003vq}
R.~R. Caldwell, M.~Kamionkowski, and N.~N. Weinberg, {\em Phantom energy and
  cosmic doomsday}, Phys. Rev. Lett. {\bf 91} (2003) 071301 
[\href{http://arxiv.org/abs/astro-ph/0302506v1}{astro-ph/0302506v1}].

\bibitem{ArkaniHamed:2003uy}
N.~Arkani-Hamed, H.-C. Cheng, M.~A. Luty, and S.~Mukohyama, {\em Ghost
  condensation and a consistent infrared modification of gravity}, JHEP {\bf 05} (2004) 074 
[\href{http://arxiv.org/abs/hep-th/0312099v1}{hep-th/0312099v1}].


\bibitem{Planck2013}
P. A. R. Ade et al., {\em Planck 2015 results. XIII. Cosmological parameters}, (2015) 
[\href{http://arxiv.org/abs/1502.01589v2}{arXiv:1502.01589v2}].

\bibitem{JoaoM}
  J.~Morais, M.~Bouhmadi-L\'opez and S.~Capozziello,
  {\em Can $f(R)$ gravity contribute to (dark) radiation?},
  JCAP {\bf 1509} no.09  (2015) 041 
[\href{http://arxiv.org/abs/1507.02623v3}{arXiv:1507.02623v3}].

\bibitem{Bouhmadi-Lopez:2014cca}
M.~Bouhmadi-L\'opez, A.~Errahmani, P.~Mart\'in-Moruno, T.~Ouali, and
  Y.~Tavakoli, {\em The little sibling of the big rip singularity},
Int. J. Mod. Phys. D {\bf 24} no.~10 (2015) 1550078  
[\href{http://arxiv.org/abs/1407.2446v2}{arXiv:1407.2446v2}].

\bibitem{Imanol}
  I.~Albarran, M.~Bouhmadi-L\'opez, F.~Cabral and P.~Mart\'in-Moruno,
  {\em The quantum realm of the "Little Sibling" of the Big Rip singularity},
  JCAP {\bf 1511} no.11  (2015) 044 
[\href{http://arxiv.org/abs/1509.07398v1}{arXiv:1509.07398v1}].
{
\bibitem{BCP}
O.~Bertolami, P.~Carrilho and J.~P\'aramos,
{\em Two-scalar-field model for the interaction of dark energy and dark matter},
Phys. Rev. D {\bf 86}, 103522 (2012);
[\href{https://arxiv.org/abs/1206.2589}{arXiv:1206.2589}].
\bibitem{BLMZ}
M.~Bouhmadi-L\'opez, J.~Morais and A.~Zhuk,
{\em The late Universe with non-linear interaction in the dark sector: the coincidence problem};
[\href{https://arxiv.org/abs/1603.06983}{arXiv:1603.06983}].
}
\end{thebibliography}
\end{document}